\RequirePackage{filecontents}

\RequirePackage{snapshot}
\documentclass[aps,pra,reprint,showpacs,secnumarabic,superscriptaddress]{preprint}

\renewcommand\documentclass[2][]{}
\documentclass[aps,pra,twocolumn,showpacs,floatfix,superscriptaddress]{revtex4-1}

\usepackage[T1]{fontenc}
\usepackage{amsmath}
\usepackage{txfonts}
\usepackage{tgtermes}
\usepackage{qtxmath}
\renewcommand*{\vec}[1]{\boldsymbol{#1}}
\usepackage{microtype}
\usepackage{graphicx}
\graphicspath{{figures/}}
\usepackage{color}
\usepackage[breaklinks=true,
pdfauthor={Heiko Bauke, Sven Ahrens, Christoph H. Keitel, and Rainer Grobe},
pdftitle={Relativistic spin operators in various electromagnetic
  environments}]{hyperref}
\usepackage{braket}
\usepackage{ifthen}

\newcommand*{\op}[1]{\hat{#1}}
\newcommand*{\opHo}{\op{H}_0}
\newcommand*{\opHC}{\op{H}_C}
\newcommand*{\opp}{\op{\vec p}}
\newcommand*{\oppo}{\op{p}_0}
\newcommand*{\opsigma}[1][]{\ifthenelse{\equal{#1}{}}{\vec{\op{\sigma}}}{\op{\sigma}_{#1}}}
\newcommand*{\opSigma}[1][]{\ifthenelse{\equal{#1}{}}{\vec{\op{\Sigma}}}{\op{\Sigma}_{#1}}}
\newcommand*{\opS}[1][]{\ifthenelse{\equal{#1}{}}{\op{\vec{S}}}{\op{S}_{#1}}}
\newcommand*{\opL}[1][]{\ifthenelse{\equal{#1}{}}{\op{\vec{L}}}{\op{L}_{#1}}}
\newcommand*{\opJ}[1][]{\ifthenelse{\equal{#1}{}}{\op{\vec{J}}}{\op{J}_{#1}}}
\newcommand*{\opN}[1][]{\ifthenelse{\equal{#1}{}}{\op{\vec{N}}}{\op{N}_{#1}}}
\newcommand*{\opW}[1][]{\ifthenelse{\equal{#1}{}}{\op{\vec{W}}}{\op{W}_{#1}}}
\newcommand*{\opSP}[1][]{\ifthenelse{\equal{#1}{}}{\vec{\op{S}}_\mathrm{P}}{\op{S}_{\mathrm{P},#1}}}
\newcommand*{\opSPnr}[1][]{\ifthenelse{\equal{#1}{}}{\vec{\op{S}}_\mathrm{P,nr}}{\op{S}_{\mathrm{P,nr},#1}}}
\newcommand*{\opSFW}[1][]{\ifthenelse{\equal{#1}{}}{\vec{\op{S}}_\mathrm{FW}}{\op{S}_{\mathrm{FW},#1}}}
\newcommand*{\opSCz}[1][]{\ifthenelse{\equal{#1}{}}{\vec{\op{S}}_\mathrm{Cz}}{\op{S}_{\mathrm{Cz},#1}}}
\newcommand*{\opSF}[1][]{\ifthenelse{\equal{#1}{}}{\vec{\op{S}}_\mathrm{F}}{\op{S}_{\mathrm{F},#1}}}
\newcommand*{\opSCh}[1][]{\ifthenelse{\equal{#1}{}}{\vec{\op{S}}_\mathrm{Ch}}{\op{S}_{\mathrm{Ch},#1}}}
\newcommand*{\opSPr}[1][]{\ifthenelse{\equal{#1}{}}{\vec{\op{S}}_\mathrm{Pr}}{\op{S}_{\mathrm{Pr},#1}}}
\newcommand*{\opSFG}[1][]{\ifthenelse{\equal{#1}{}}{\op{\vec{S}}_\mathrm{FG}}{\op{S}_{\mathrm{FG},#1}}}

\newcommand*{\F}[3][\pm]{\phi^{#1}_{#2,#3}(\vec r, t)}

\newcommand*{\I}{\mathrm{i}}
\newcommand*{\D}{\mathrm{d}}
\newcommand*{\e}{\mathrm{e}}
\newcommand*{\commut}[2]{%
  \mathchoice{\left[#1,#2\right]}{[#1,#2]}{[#1,#2]}{[#1,#2]}%
}
\newcommand*{\trans}[1]{#1^{\mathsf{T}}}
\newcommand*{\htrans}[1]{#1^{\dagger}}
\newcommand*{\GAMMA}{\mathrm{\Gamma}}

\DeclareMathOperator{\spec}{spec}
\DeclareMathOperator{\RE}{Re}

\begin{document}

\clubpenalty = 2500
\widowpenalty = 5000
\displaywidowpenalty = 5000

\title{Relativistic spin operators in various electromagnetic
  environments}

\author{Heiko Bauke}
\affiliation{Max-Planck-Institut f{\"u}r Kernphysik, Saupfercheckweg~1,
  69117~Heidelberg, Germany} 
\author{Sven Ahrens}
\affiliation{Max-Planck-Institut f{\"u}r Kernphysik, Saupfercheckweg~1,
  69117~Heidelberg, Germany} 
\affiliation{Intense Laser Physics Theory Unit and Department of Physics, Illinois State University, Normal, Illinois 61790-4560, USA}
\author{Christoph H. Keitel}
\affiliation{Max-Planck-Institut f{\"u}r Kernphysik, Saupfercheckweg~1,
  69117~Heidelberg, Germany} 
\author{Rainer Grobe}
\affiliation{Max-Planck-Institut f{\"u}r Kernphysik, Saupfercheckweg~1,
  69117~Heidelberg, Germany} 
\affiliation{Intense Laser Physics Theory Unit and Department of Physics, Illinois State University, Normal, Illinois 61790-4560, USA}

\date{\today}

\begin{abstract}
  Different operators have been suggested in the literature to describe
  the electron's spin degree of freedom within the relativistic Dirac
  theory.  We compare concrete predictions of the various proposed
  relativistic spin operators in different physical situations.  In
  particular, we investigate the so-called Pauli, Foldy-Wouthuysen,
  Czachor, Frenkel, Chakrabarti, Pryce, and Fradkin-Good spin operators.
  We demonstrate that when a quantum system interacts with
  electromagnetic potentials the various spin operators predict
  different expectation values.  This is explicitly illustrated for the
  scattering dynamics at a potential step and in a standing laser field
  and also for energy eigenstates of hydrogenic ions.  Therefore, one
  may distinguish between the proposed relativistic spin operators
  experimentally.
\end{abstract}

% 03.65.Pm      Relativistic wave equations 
% 31.15.aj      Relativistic corrections, spin-orbit effects, fine structure; hyperfine structure
% 31.30.J-      Relativistic and quantum electrodynamic (QED) effects in atoms, molecules, and ions 

\maketitle

\section{Introduction}
\label{sec:intro}

Elementary particles such as the electron carry some internal
angular-momentum-like degree of freedom that is called spin. It is
well understood that angular momentum is intrinsically tied to the
group-theoretic structure of (relativistic) quantum mechanics
\cite{Wigner:1939:Unitary_Representations}.  The understanding of the
physical nature of the spin, however, is still incomplete
\cite{Morrison:2007:Spin,Nikolic:2007:Myths_Facts,Giulini:2008:spin}.
Historically, the concept of spin was introduced in order to explain
some experimental findings such as the emission spectra of alkali
metals and the Stern-Gerlach experiment.  A direct measuring of the
spin (or more precisely the electron's magnetic moment), however, was
missing until the pioneering work by Dehmelt
\cite{Dehmelt:1990:Structure}.  Pauli and Bohr even claimed that the
spin of free electrons was impossible to measure for fundamental
reasons \cite{Mott:1929:Scattering}.  Recent renewed interest in
fundamental aspects of the spin
\cite{Caban:Rembielinski:etal:2013:Spin_operator} arose, for example,
from high-precision measurement experiments for the electron's
magnetic moment
\cite{Hanneke:etal:2008:New_Measurement,Neumann:etal:2010:Single-Shot_Readout,Buckley:2010:Spin-Light_Coherence,Close:etal:2011:Spin_State_Amplification,Sturm:etal:2011:g_factor,DiSciacca:2012:Direct_Measurement},
the growing field of (relativistic) quantum information theory
\cite{Czachor:1997:relativistic_EPR,Peres:Terno:2004:QI_RT,Kim:etal:2005:Bell,Wiesendanger:2009:Spin_mapping,Caban:etal:2010:Polarization_correlations,Friis:etal:2010,Petersson:2010:Circuit_qed,Saldanha:Vedral:2012:Wigner},
quantum spintronics \cite{Awschalom:etal:2013:Quantum_Spintronics},
spin effects in graphene
\cite{Abanin:2011:Giant_Spin-Hall_Effect,Mecklenburg:Regan:2011:Spin,Guttinger:2010:Spin_States},
and light-matter interactions at relativistic intensities
\cite{Walser:Urbach:etal:2002:Spin_and,DiPiazza:etal:2012:high-intensity,Ahrens:etal:2012:spin_dynamics,Klaiber:etal:2014:Spin_dynamics,Ahrens:etal:2014:Electron-spin_dynamics_induced_by_photon_spins}.

Although the spin is regarded as a fundamental property of the
electron, there is no universally accepted spin operator.  In fact,
one can find in the literature several proposals of different spin
operators for the Dirac theory
\cite{Caban:Rembielinski:etal:2013:Spin_operator}.  These operators
are often motivated by abstract group-theoretic considerations rather
than by experimental evidence.  In our view, there are very few works
that consider specific experimental schemes and predict concrete
expectation values for spin observables in a relativistic setting.
Such predictions, however, are required to judge which of the proposed
relativistic generalizations of the spin or (equivalently) of the
position operators are best suited to describe experimental
observations. For example, a paper by Czachor
\cite{Czachor:1997:relativistic_EPR} proposed to use an
Einstein-Podolsky-Rosen type of experiment and the associated degree
of violation of the Bell inequality to test various relativistic
concepts, such as the relativistic position operator. This work also
predicts that the center of mass and the center of charge might not
necessarily agree for a relativistic electron leading to possible
implications for quantum cryptography.  Another example is the work
\cite{Choi:etal:2011:Relativistic_effects} by Choi
\textit{et~al.}, who studied spin entanglement of massive Dirac
particles.

In this work we examine seven proposals for the relativistic spin
operator, which we tentatively call here the Pauli, the
Foldy-Wouthuysen, the Czachor, the Frenkel, the Chakrabarti, the
Fradkin-Good, and the Pryce spin operators.  Our aim is to investigate
and to compare their mathematical properties and to analyze how
different definitions of relativistic spin operators may lead to
different predictions for spin expectation values in various
experimental setups.  The seven spin operators discussed in this work
share the same nonrelativistic limit, obtained by assuming that the
particle's kinematic momentum is small compared to $m_0c$, with $m_0$
denoting the particle's rest mass and $c$ the speed of light.  Thus,
any differences in the spin's properties are purely relativistic
effects and require most likely accelerated particles.  While several
works have tried to relate the different functional forms of these
operators to each other, a study that shows how the predictions depend
on the choice of the relativistic spin operator for an electron whose
dynamical evolution is governed by external electromagnetic fields is
lacking.  This requires a concrete computational analysis yielding
concrete predictions about expectation values that can be directly
compared with experimental results.  Using numerical wave function
solutions to the time-dependent Dirac equation, we evaluate and
compare the various predictions that result from different
relativistic spin operators.  In this way we aim to build a bridge
between theoretical considerations and experiment.

A relativistic spin operator may be introduced by splitting the
undisputed total angular momentum operator $\opJ$ into an external
part $\opL$ and an internal part $\opS$ commonly referred to as the
orbital angular momentum and the spin, viz. $\opJ=\opL+\opS$.  Because
the orbital angular momentum is related to the position operator
$\vec{\op{r}}$ and the momentum operator $\opp=-\I\vec\nabla$ (units
are used in this paper for which $\hbar=1$) via
$\opL=\vec{\op{r}}\times\opp$, different definitions of the spin
operator $\vec{\op S}$ imply different relativistic position operators
$\vec{\op r}$.  The latter would be difficult to discriminate
experimentally as it couples only to a gravitational field, while the
spin couples also to the more easily controllable magnetic field.  The
rather fundamental question which mathematical operators actually
correspond to experimentally measured observables has become more
relevant as laser-particle experiments have entered the regime in
which the particle must be described in a fully relativistic framework
\cite{Salamin:etal:2006:Relativistic,DiPiazza:etal:2012:high-intensity}.

This paper is organized as follows.  In Sec.~\ref{sec:Dirac} we
briefly review the Dirac equation and introduce some notation that
will be utilized in Sec.~\ref{sec:spins}, where the seven spin
operators are defined and their mathematical properties are
analyzed.  % In
% Sec.~\ref{sec:eigen} we construct simultaneous eigenfunctions of the
% free Dirac Hamiltonian and the various spin operators and relate them to
% each other.  
The expectation values of the spin operators are evaluated for
relativistic scattering dynamics in Sec.~\ref{sec:scattering} and for
the bound states of hydrogenic ions in Sec.~\ref{sec:hydrogen}.  We
formulate our conclusions in Sec.~\ref{sec:conclusions}.

\section{The Dirac equation}
\label{sec:Dirac}

A Lorentz invariant quantum mechanical description of the motion of an
electron is given by the time-dependent Dirac equation.  For a particle
of rest mass $m_0$ and charge $q$ moving in the electromagnetic
potentials $\phi(\vec r, t)$ and $\vec A(\vec r, t)$ it is given by
\begin{multline}
  \label{eq:Dirac}
  \I\frac{\partial \Psi(\vec r, t)}{\partial t} =
  \op H\Psi(\vec r, t) = \\
  \left(
    c\vec\alpha\cdot\left(\opp- q \vec A(\vec r, t) \right) + 
    q \phi(\vec r, t) + m_0c^2\beta
  \right)\Psi(\vec r, t)\,,
\end{multline}
with the matrices $\vec\alpha=\trans{(\alpha_1, \alpha_2, \alpha_3)}$
and $\beta$. These $4\times4$ matrices obey the algebra
\begin{equation}
  \label{eq:Dirac_algebra}
  \alpha_i^2 = \beta^2 = 1\,,\quad
  \alpha_i\alpha_k + \alpha_k\alpha_i = 2\delta_{i,k}\,,\quad
  \alpha_i\beta + \beta\alpha_i = 0\,.
\end{equation}
To briefly discuss our notation and abbreviations, we use the Dirac
representation for the matrices $\alpha_i$ and $\beta$ such that
\begin{equation}
  \alpha_i =
  \begin{pmatrix}
    0 & \sigma_i \\ \sigma_i & 0
  \end{pmatrix}\,,
  \qquad
  \beta =
  \begin{pmatrix}
    \mathbb{I}_2 & 0 \\ 0 & -\mathbb{I}_2
  \end{pmatrix}\,,
\end{equation}
where the three $2\times2$ Pauli matrices $\vec\sigma=\trans{(\sigma_1,
  \sigma_2, \sigma_3)}$ are given by
\begin{equation}
  \sigma_1 = \begin{pmatrix} 0 & 1 \\ 1 & 0 \end{pmatrix} \,,\quad
  \sigma_2 = \begin{pmatrix} 0 & -\I \\ \I & 0 \end{pmatrix} \,,\quad
  \sigma_3 = \begin{pmatrix} 1 & 0 \\ 0 & -1 \end{pmatrix} 
\end{equation}
and $\mathbb{I}_2$ denotes the $2\times2$ identity matrix.  The
free-particle Dirac Hamiltonian with $\vec A(\vec r, t)=0$ and
$\phi(\vec r, t)=0$ will be denoted by $\opHo$.  The doubly degenerate
spectrum of the free Dirac Hamiltonian is given by $\spec(\op H_0)=\pm
cp_0(\vec p)$, where $p_0(\vec p)$ is the scaled positive energy
$p_0(\vec p)=(m_0^2c^2 + \vec p^2)^{1/2}$ for the momentum vector
$\vec p$.  We will also use the operator $\op p_0$ to denote
\begin{equation}
  \op p_0=\sqrt{ m_0^2c^2 + \opp^2} \,.
\end{equation}
For a given momentum eigenvalue $\vec p$, the associated eigenvectors
for the positive and negative energies $\pm p_0(\vec p)c$ can be
chosen as
\begin{subequations}
  \label{eq:pos_neg_free}
  \begin{align}
    \label{eq:pos_free}
    \F[+]{\vec\chi}{\vec p} & = 
    \vec u_{\vec \chi,\vec p}\e^{\I(\vec p\cdot\vec r - cp_0(\vec p) t)}\,,\\
    \label{eq:neg_free}
    \F[-]{\vec\chi}{\vec p} & = 
    \vec v_{\vec \chi,\vec p}\e^{\I(\vec p\cdot\vec r + cp_0(\vec p) t)}\,,
  \end{align}
\end{subequations}
where we have introduced the vectors
\begin{subequations}
  \begin{align}
    \vec u_{\vec \chi,\vec p} & = \sqrt{\frac{m_0c + p_0(\vec
        p)}{2p_0(\vec p)}}
    \begin{pmatrix}
      \vec\chi \\[1ex] \dfrac{\vec\sigma\cdot\vec p}{m_0c + p_0(\vec p)}
      \vec\chi
    \end{pmatrix}\,, \\
    \vec v_{\vec \chi,\vec p} & = \sqrt{\frac{m_0c + p_0(\vec
        p)}{2p_0(\vec p)}}
    \begin{pmatrix}
      -\dfrac{\vec\sigma\cdot\vec p}{m_0c + p_0(\vec p)} \vec\chi \\
      \vec\chi
    \end{pmatrix}\,.
  \end{align}
\end{subequations}
The quantity $\vec\chi$ denotes an arbitrary complex two-component
vector with $\htrans{\vec\chi}\cdot\vec\chi=1$.  Note that while
(\ref{eq:pos_free}) corresponds to states that travel in the direction
given by the vector $\vec p$, the states given by (\ref{eq:neg_free})
travel in the opposite direction of $\vec p$. The two fold degenerate
eigenspace of $\opHo$ for each eigenvalue $cp_0$ can be spanned by the
two (mutually orthogonal) eigenfunctions $\F[+]{\vec\chi}{\vec p}$ and
$\F[+]{\vec\chi_\perp}{\vec p}$, where the normalized vector
$\vec\chi_{\perp}$ is orthogonal to $\vec\chi$.

Obviously, any superposition of the two functions $\F[+]{\vec\chi}{\vec
  p}$ and $\F[+]{\vec\chi_\perp}{\vec p}$ is also an energy eigenstate.
Analogous statements hold for the negative-energy eigenstates.  The
functions $\F{\vec\chi}{\vec p}$ and $\F{\vec\chi_\perp}{\vec p}$ form a
basis, thus each wave packet can be written as a superposition of
$\F{\vec\chi}{\vec p}$ and $\F{\vec\chi_\perp}{\vec p}$.  In the course
of our presentation, it will be useful to introduce the energy subspace
operators
\begin{equation}
  \label{eq:Lambda_pm}
  \op{\Lambda}^\pm = 
  \frac{1}{2} \left(1\pm\frac{\opHo}{c\oppo}\right)
\end{equation}
that single out positive- and negative-energy contributions,
respectively, from an arbitrary superposition.

\section{Seven variations on spin}
\label{sec:spins}

\begin{table*}
  \centering
  \caption{Brief summary of the proposed spin operators' definitions
    and their mathematical properties. The table indicates from left to
    right the definition of the various spin operators, if they commute
    with the free Dirac Hamiltonian, if they obey the angular momentum
    algebra, if eigenvalues are $\pm1/2$, and if the operators are
    related to the Pauli spin operator via an orthogonal transformation.}
  \label{tab:properties}
  \begin{ruledtabular}
    \begin{tabular}{@{}lccccc@{}}
      %operator name & 
      Definition & $\opS=\htrans{\opS}$ & $\commut{\opHo}{\opS}=0$ &
      $\commut{\opS[i]}{\opS[j]}=\I\varepsilon_{i,j,k}\opS[k]$ &
      Eigenvalues equal to $\pm1/2$ & $\opS = \op{T}\opSP\op{T}^{-1}$ \\
      \colrule
      \rule{0pt}{4ex}%
      %Pauli spin operator
      %\cite{Hill:Landshoff:1938,Dirac:1958:principles,Dirac:1971:Positive-Energy,Ohanian:1986:spin,Lifshitz:1996:QED,Feynman:1998:QED} & 
      $\opSP = \dfrac 12 \opSigma$ &
      yes & no & yes & yes & --- \\[1.75ex]
      %Foldy-Wouthuysen spin operator 
      %\cite{Foldy:Wouthuysen:1950:Non-Relativistic_Limit,deVries:1970:FW_Transformation,Costella:McKellar:1995:Foldy-Wouthuysen,Schweber:2005:qft,Caban:etal:2013:A_spin_observable} & 
      $ \opSFW = \dfrac{1}{2}\opSigma + 
      \dfrac{\I\beta}{2\oppo}\op{\vec{p}}\times\vec\alpha -
      \dfrac{\op{\vec{p}}\times(\opSigma\times\op{\vec{p}})}{2\oppo(\oppo+m_0c)}$ & 
      yes & yes & yes & yes & yes \\[1.25ex]
      % Czachor spin operator \cite{Czachor:1997:relativistic_EPR} & 
      $\opSCz % = \dfrac{1}{2}\left(
      %   \op{\Lambda}^+\opSigma\op{\Lambda}^+ + 
      %   \op{\Lambda}^-\opSigma\op{\Lambda}^-
      % \right) 
      = \dfrac{m_0^2c^2}{2\oppo^2}\opSigma + 
      \dfrac{\I m_0c\beta}{2\oppo^2}\op{\vec{p}}\times\vec\alpha +
      \dfrac{\op{\vec{p}}\cdot\opSigma}{2\oppo^2}\op{\vec{p}}$ & 
      yes & yes & no & no & no \\[1.75ex]
      % Frenkel spin operator 
      % \cite{Hilgevoord:Wouthuysen:1963:Dirac_spin,Wightman:1960,Bargmann:etal:1935:Precession} & 
      $\opSF = \dfrac{1}{2}\opSigma +
      \dfrac{\I\beta}{2m_0c}\op{\vec{p}}\times\vec\alpha$ & 
      yes & yes & no & no & no \\[1.75ex]
      % Chakrabarti spin operator
      % \cite{Chakrabarti:1963:Canonical_Form,Gursey:1965:Equivalent_formulations,Gursey:1965,Gursey:1965:Equivalent_formulations,Choi:2012:Relativistic_spin_operator} & 
      $ \opSCh = 
      \dfrac{1}{2}\opSigma + 
      \dfrac{\I}{2m_0c}\vec\alpha\times\op{\vec{p}} +
      \dfrac{\op{\vec{p}}\times(\opSigma\times\op{\vec{p}})}{2m_0c(m_0c+\oppo)}$ &
      no & no & yes & yes & yes \\[1.75ex]
      % Pryce spin operator
      % \cite{Pryce:1935:Commuting_Co-Ordinates,Pryce:1948:Mass-Centre,Macfarlane:1963:Relativistic_Two_Particle,Berg:1965:Position_Spin,Ryder:1999:Relativistic_Spin} & 
      $ \opSPr = 
      % \dfrac{1}{2}\beta\opSigma + 
      % \dfrac{\I c\alpha_3\alpha_2\alpha_1(\beta+1)}{2(\opHo+m_0c^2)} \op{\vec{p}} =
      % \dfrac{1}{2}\beta\opSigma + 
      % \dfrac{\I\alpha_3\alpha_2\alpha_1(\beta+1)\vec{\alpha}\cdot\op{\vec{p}}}{2\op{\vec{p}}^2}\op{\vec{p}}
      \dfrac{1}{2}\beta\opSigma + 
      \dfrac{1}{2}\opSigma\cdot\opp(1-\beta)\dfrac{\opp}{\opp^2}$ & 
      yes & yes & yes & yes & yes \\[1.75ex]
      $ \opSFG = 
      \dfrac{1}{2}\beta\opSigma + 
      \dfrac{1}{2}\opSigma\cdot\opp\left(\dfrac{\opHo}{c\oppo}-\beta\right)\dfrac{\opp}{\opp^2}$ & 
      yes & yes & no & yes & no
    \end{tabular}
  \end{ruledtabular}
\end{table*}

In this section we will define seven different spin operators referred
to as Pauli, Foldy-Wouthuysen, Czachor, Frenkel, Chakrabarti, Pryce,
and Fradkin-Good spin operators.  Each of these operators is
characterized by a triplet $\opS=\trans{(\opS[1],\opS[2],\opS[3])}$.
For simplicity, we will also denote the spin component in a given
$\vec n$ direction by $\opS[\vec n]$ defined as $\opS[\vec n]=\vec
n\cdot\opS$.  For some calculations it will be beneficial to
parametrize the vector $\vec n$ as $\vec
n=\trans{(\sin\vartheta\cos\varphi,
  \sin\vartheta\sin\varphi,\cos\vartheta)}$ and to define the two
orthogonal vectors
\begin{equation}
  \label{eq:ch_defi}
  \vec\chi_\uparrow = 
  \begin{pmatrix}
    \cos(\vartheta/2) \\
    \sin(\vartheta/2)\,\e^{\I\varphi}
  \end{pmatrix}\,,\quad
  \vec\chi_\downarrow = 
  \begin{pmatrix}
    -\sin(\vartheta/2)\,\e^{-\I\varphi} \\
    \cos(\vartheta/2) 
  \end{pmatrix}\,,
\end{equation}
which are the eigenvectors of $\vec n\cdot\vec\sigma$.  Furthermore, we
define the triplet of operators $\opSigma=\trans{ (\opSigma[1],
  \opSigma[2], \opSigma[3])}$ via
\begin{equation}
  \op\Sigma_i = -\I\alpha_j \alpha_k
\end{equation}
with $(i,j,k)$ being a cyclic permutation of $(1,2,3)$.  Its individual
components fulfill the usual angular momentum commutator relationship
$\commut{\opSigma[i]}{\opSigma[j]}=2\I\varepsilon_{i,j,k}\opSigma[k]$
with the Levi-Civita symbol $\varepsilon_{i,j,k}$.  This operator is
normalized to $\htrans{\vec{\opSigma}}\cdot\vec{\opSigma}=3$ and its
components have the doubly degenerate eigenvalues $\pm1$.  The standard
representation of $\opSigma$ is given by
\begin{equation}
  \opSigma[i] =
  \begin{pmatrix}
    \sigma_i & 0 \\ 0 & \sigma_i 
  \end{pmatrix}\,.
\end{equation}
Various spin operators can be defined in terms of the Pauli-Lubanski
vector $\opW$ and the related scalar operator $\op{W}_0$.  Introducing
the generator of the Lorentz boosts
\begin{equation}
  \opN = \frac{1}{2c^2}(\vec{r}\opHo+\opHo\vec{r})\,,
\end{equation}
$\opW$ and $\op{W}_0$ are defined as
\begin{gather}
  \label{eq:Pauli-Lubanski_W}
  \opW = \frac{1}{c}\opHo\opJ + c\opp\times\opN 
  = \frac{1}{4c}(\opHo\opSigma+\opSigma\opHo)\,, \\
  \label{eq:Pauli-Lubanski_W0}
  \op{W}_0 = \opp\cdot\opJ
  = \frac{1}{2}\opp\cdot\opSigma\,.
\end{gather}
With these definitions we are prepared now to summarize briefly the
proposed spin operators, to give their explicit expressions, and to
discuss some of their properties.  An overview of the proposed spin
operators is also given in Table~\ref{tab:properties}.

\subsection{Pauli spin operator}

The Pauli spin operator
\cite{Hill:Landshoff:1938,Dirac:1958:principles,Dirac:1971:Positive-Energy,Ohanian:1986:spin,Lifshitz:1996:QED,Feynman:1998:QED}
is a direct generalization of the spin operator of nonrelativistic
quantum mechanics.  Expressing the total angular momentum operator
$\vec{\op J}$ as $\vec{\op J}=\vec r\times \opp + \opSigma/2$, it
appears quite natural to identify $\vec r\times \opp$ as the
orbital angular momentum and to define
\begin{equation}
  \label{eq:SP}
  \opSP = \frac 12 \opSigma
\end{equation}
as the relativistic Pauli spin operator.  In many standard textbooks
on relativistic quantum dynamics
\cite{Dirac:1958:principles,Lifshitz:1996:QED,Feynman:1998:QED} this
operator is considered as the relativistic spin operator.  The energy
shift for a hydrogenic ground state $\psi_{\uparrow}$ (see
Sec.~\ref{sec:hydrogen}) that is exposed to a weak homogeneous
magnetic field $\vec B=\trans{(0,0,B)}$ (anomalous Zeeman effect)
relative to the field-free case is, with the atomic number $Z$
\cite{Margenau:1940:Relativistic_Magnetic_Moment,Mott:Massey:1965:Atomic_Collisions},
\begin{equation}
  \frac{Bq}{m_0}\frac{1}{6}\left(1+2\sqrt{1-Z^2\alpha_\mathrm{el}^2}\right) =
  \frac{Bq}{m_0}\braket{\psi_{\uparrow} | \opSP[3] | \psi_{\uparrow} }
\end{equation}
(with $\alpha_\mathrm{el}$ denoting the fine-structure constant),
which is often brought up as an argument for $\opSP$ representing the
relativistic spin \cite{Dirac:1958:principles}.

The components of the Pauli spin operator are generators of the SU(2)
algebra and fulfill the angular momentum algebra
\begin{equation}
  \commut{\opSP[i]}{\opSP[j]}=\I\varepsilon_{i,j,k}\opSP[k]\,;
\end{equation}
the total squared length is $\opSP^2=3/4$.  The degenerate eigenvalues
$s_\mathrm{P}$ and the normalized orthogonal eigenvectors $\vec
s_\mathrm{P}$ of $\opSP[\vec n]$ are given by
\begin{align}
  \label{eq:Pauli_s}
  \nonumber
  s_{\mathrm{P}\uparrow} & = \frac 12 \,: & 
  \vec s_{\mathrm{P}\uparrow,1} & =
  \vec u_{\vec\chi_\uparrow, 0}\e^{\I\vec p\cdot\vec r} \,, &
  \vec s_{\mathrm{P}\uparrow,2} & =
  \vec v_{\vec\chi_\uparrow, 0}\e^{\I\vec p\cdot\vec r} \, \\
  s_{\mathrm{P}\uparrow} & = -\frac 12 \,: & 
  \vec s_{\mathrm{P}\downarrow,1} & =
  \vec u_{\vec\chi_\downarrow, 0}\e^{\I\vec p\cdot\vec r} \,, &
  \vec s_{\mathrm{P}\downarrow,2} & =
  \vec v_{\vec\chi_\downarrow, 0}\e^{\I\vec p\cdot\vec r} \,,
\end{align}
with $\vec\chi_\uparrow$ and $\vec\chi_\downarrow$ as defined in
\eqref{eq:ch_defi}, which are also the eigenvectors of the
nonrelativistic Pauli spin operator $ \opSPnr = {\opsigma}/{2}$.  The
relativistic Pauli spin operator does not commute with the free
Hamiltonian,
\begin{equation}
  \commut{\opHo}{\opSP[\vec n]} = 
  \I c\vec n(\op{\vec p}\times\vec\alpha) \,.
\end{equation}
As a consequence of this nonvanishing commutator, even for a free
particle the expectation value of $\opSP[\vec n]$ can evolve
nontrivially in time, leading, for example, to the zitterbewegung
\cite{Thaller:1992:Dirac,Braun:etal:1999:Numerical_approach,Krekora:etal:2004:localization}
of the Pauli spin, if the quantum state is a superposition of states
of positive- and negative-energy solutions of the free Dirac
Hamiltonian.  This is often considered an undesirable feature for a
relativistic spin operator because an intrinsic observable should be
constant when no forces act.

\subsection{Foldy-Wouthuysen spin operator}

A second definition of the spin is based on the Foldy-Wouthuysen
transformation
\cite{Foldy:Wouthuysen:1950:Non-Relativistic_Limit,deVries:1970:FW_Transformation,Costella:McKellar:1995:Foldy-Wouthuysen,Schweber:2005:qft,Caban:etal:2013:A_spin_observable},
which is a unitary transformation $\op{T}_\mathrm{FW}$ that turns the
Dirac equation \eqref{eq:Dirac} into block-diagonal form, reducing
positive- and negative-energy states to two-component wave functions.
For the free-particle Hamiltonian $\opHo$ Foldy and Wouthuysen showed
that
\begin{equation}
  \op{T}_\mathrm{FW}^{-1}\opHo\op{T}_\mathrm{FW} = c\beta\oppo\,,
\end{equation}
with
\begin{equation}
  \label{eq:UFW}
  \op{T}_\mathrm{FW} = 
  \frac{\oppo + m_0c -\beta\vec{\alpha}\cdot\opp}
  {\sqrt{2\oppo(\oppo + m_0c)}}\,.
\end{equation}
Furthermore, Foldy and Wouthuysen postulated that the spin operator in
the transformed representation is $\opSP$ indeed, leading to the
Foldy-Wouthuysen spin operator
\begin{equation}
  \opSFW =
  \op{T}_\mathrm{FW}\opSP\op{T}_\mathrm{FW}^{-1}
\end{equation}
or more explicitly 
\begin{equation}
  \label{eq:SFW}
  \opSFW = \frac{1}{2}\opSigma + 
  \frac{\I\beta}{2\oppo}\opp\times\vec\alpha -
  \frac{\opp\times(\opSigma\times\opp)}{2\oppo(\oppo+m_0c)}
  \,.
\end{equation}
Two years before the celebrated Foldy-Wouthuysen paper
\cite{Foldy:Wouthuysen:1950:Non-Relativistic_Limit} the equivalent
expression
\begin{equation}
  \label{eq:SFW_bar}
  \opSFW = 
  \frac{1}{2c\oppo}\left(
    m_0c^2\opSigma - \I c\beta\vec\alpha\times\opp +
    \frac{c^2\opp\cdot\opSigma}{c\oppo+m_0c^2}\opp
  \right)
\end{equation}
for the Foldy-Wouthuysen spin operator was given by Pryce in
\cite{Pryce:1948:Mass-Centre}.  In this publication it was also shown
that this spin operator is closely related to the Czachor and the
Frenkel spin operators via the associated position operators.  A
further representation of the Foldy-Wouthuysen spin operator
\eqref{eq:UFW} can be written in terms of the Pauli-Lubandski vector
\cite{Caban:etal:2013:A_spin_observable}
\begin{equation}
   \label{eq:SFW_bar_bar}
   \opSFW = \frac{1}{m_0c}\left(
     \frac{c\oppo}{\opHo}\opW - \frac{\op{W}_0}{\oppo+m_0c}\opp
   \right)\,.
\end{equation}
A further equivalent expression that is sometimes given in the
literature is given by
\cite{Caban:Rembielinski:etal:2013:Spin_operator}
\begin{equation}
  \label{eq:SFW_bar_bar_bar}
  \opSFW = \frac{\oppo}{2m_0c}\opSigma - 
  \frac{\opp\cdot\opSigma}{2m_0c(m_0c+\oppo)}\opp - 
  \opp\times\vec\alpha\frac{\I\opHo}{2m_0c^2\oppo}\,.
\end{equation}
 
As $\opSFW$ is unitarily equivalent to $\opSP$, its components fulfill
the same commutator relationships.  From $\commut{c\beta\oppo}{\opSP}=0$
it follows $\commut{\opHo}{\opSFW}=0$. Thus, the Foldy-Wouthuysen spin
operator is conserved for free particles and $\opSFW[\vec n]$ and
$\opHo$ have a common set of eigenvectors.  The degenerate eigenvalues
$s_\mathrm{FW}$ and the normalized orthogonal eigenvectors $\vec
s_\mathrm{FW}$ of $\opSFW[\vec n]$ and $\opHo$ are given by
\begin{align}
  \label{eq:SFW_eigen}
  \nonumber
  s_{\mathrm{FW}\uparrow} & = \frac 12 \,: & 
  \vec s_{\mathrm{FW}\uparrow,1} & =
  \vec u_{\vec\chi_\uparrow, \vec p}\e^{\I\vec p\cdot \vec r} \,, &
  \vec s_{\mathrm{FW}\uparrow,2} & =
  \vec v_{\vec\chi_\uparrow, \vec p}\e^{\I\vec p\cdot \vec r}\,, \\
  s_{\mathrm{FW}\downarrow} & = -\frac 12 \,: & 
  \vec s_{\mathrm{FW}\downarrow,1} & =
  \vec u_{\vec\chi_\downarrow, \vec p}\e^{\I\vec p\cdot \vec r} \,, &
  \vec s_{\mathrm{FW}\downarrow,2} & =
  \vec v_{\vec\chi_\downarrow, \vec p}\e^{\I\vec p\cdot \vec r} \,.
\end{align}
The eigenvectors $\vec s_{\mathrm{FW}\uparrow,1}$ and $\vec
s_{\mathrm{FW}\downarrow,1}$ have the positive-energy eigenvalue
$cp_0(\vec p)$, whereas $\vec s_{\mathrm{FW}\uparrow,2}$ and $\vec
s_{\mathrm{FW}\downarrow,2}$ have the negative-energy eigenvalue
$-cp_0(\vec p)$.

\subsection{Czachor spin operator}

The third spin vector has been discussed by Czachor
\cite{Czachor:1997:relativistic_EPR}, but already appeared in an
earlier work by Pryce \cite{Pryce:1948:Mass-Centre}.  It can be
defined on the basis of the spatial components of the Pauli-Lubanski
vector $\vec{\op{W}}$.  If we multiply the vector $\vec{\op{W}}$ with
the inverse of the free Dirac Hamiltonian, we can define the
Pauli-Lubanski-based spin operator
\begin{equation}
  \opSCz=\vec{\op{W}}c\opHo^{-1} \,.
\end{equation}
Using energy subspace projection operators (\ref{eq:Lambda_pm}), we can
rewrite this particular spin operator in the form
\begin{equation}
  \label{eq:SCz}
  \opSCz = \frac{1}{2}\left(
    \op{\Lambda}^+\opSigma\op{\Lambda}^+ + 
    \op{\Lambda}^-\opSigma\op{\Lambda}^-
  \right)\,.
\end{equation}
Using the projector-based representation (\ref{eq:SCz}), one can
easily see that the individual spin components cannot satisfy the
usual angular momentum commutator relationships.  For a comparison
with the other spin vectors, we rewrite $\opSCz$ also in the more
explicit form
\begin{equation}
  \label{eq:SCz2}
  \opSCz = \frac{m_0^2c^2}{2\oppo^2}\opSigma + 
  \frac{\I m_0c\beta}{2\oppo^2}\opp\times\vec\alpha +
  \frac{\opp\cdot\opSigma}{2\oppo^2}\opp\,.
\end{equation}

While the Czachor spin operator has the nice feature that it commutes
with the free Dirac Hamiltonian
\begin{equation}
  \commut{\opHo}{\opSCz[\vec n]}=0 \,,
\end{equation}
its components do not fulfill the angular momentum algebra.  In fact,
the commutator relation
\begin{equation}
  \commut{\opSCz[i]}{\opSCz[j]} = 
  \I\varepsilon_{i,j,k}\left(
    \opSCz[k] - \frac{\opSigma\cdot\op{\vec p}}{2\oppo^2}\op{p}_k
  \right)
\end{equation}
holds.  Consequently, the absolute values of the Czachor spin operator's
eigenvalues are not equal to $1/2$. The degenerate eigenvalues
$s_\mathrm{Cz}$ and the nonnormalized orthogonal eigenvectors $\vec
s_\mathrm{Cz}$ of $\opSCz[\trans{(0,0,1)}]$ are given by
\begin{align}
  \nonumber
  s_{\mathrm{Cz}\uparrow} & = \frac{p_0(\vec p_\|)}{2p_0(\vec p)} \,: & 
  \vec s_{\mathrm{Cz}\uparrow,1} & =
  \begin{pmatrix}
    p_0(\vec p_\perp)^2 + p_0(\vec p_\perp)p_0(\vec p) \\
    p_z(p_x+\I p_y) \\
    0 \\
    m_0c(p_x+\I p_y)
  \end{pmatrix}\e^{\I\vec p\cdot \vec r} \,, \\
  \nonumber
  &&
  \vec s_{\mathrm{Cz}\uparrow,2} & =
  \begin{pmatrix}
    -p_z(p_x-\I p_y) \\
    p_0(\vec p_\perp)^2 - p_0(\vec p_\perp)p_0(\vec p) \\
    m_0c(p_x-\I p_y) \\
    0 
  \end{pmatrix}\e^{\I\vec p\cdot \vec r} \,, \\
  \nonumber
  s_{\mathrm{Cz}\downarrow} & = -\frac{p_0(\vec p_\|)}{2p_0(\vec p)} \,: & 
  \vec s_{\mathrm{Cz}\downarrow,1} & =
  \begin{pmatrix}
    -p_z(p_x-\I p_y) \\
    p_0(\vec p_\perp)^2 + p_0(\vec p_\perp)p_0(\vec p) \\
    m_0c(p_x-\I p_y) \\
    0 
  \end{pmatrix}\e^{\I\vec p\cdot \vec r} \,, \\
  &&
  \vec s_{\mathrm{Cz}\downarrow,2} & =
  \begin{pmatrix}
    p_0(\vec p_\perp)^2 - p_0(\vec p_\perp)p_0(\vec p) \\
    +p_z(p_x+\I p_y) \\
    0 \\
    m_0c(p_x+\I p_y) \\
  \end{pmatrix}\e^{\I\vec p\cdot \vec r} \,,
\end{align}
where $\vec p_\perp$ and $\vec p_\|$ denote the momentum components
perpendicular and parallel to the spin orientation $\vec
n=\trans{(0,0,1)}$.  Eigenfunctions for other spin orientations may be
found by an appropriate Lorentz rotation.  The eigenvalues are functions
of the momenta that are parallel to the polarization direction of the
eigenstate. In particular, the absolute values of the Czachor spin
operator's eigenvalues are less than $1/2$.

As the Czachor spin operator and the Pauli spin operator have
different eigenvalues, the Czachor spin operator cannot be related to
the Pauli spin operator via a similarity transformation.  This means
that there is no operator $\op T_\mathrm{Cz}$ such that $\opSCz$
equals $\op T_\mathrm{Cz}\opSP\op T_\mathrm{Cz}^{-1}$.  As a
consequence of the definition \eqref{eq:SCz}, the operator identities
\begin{subequations}
  \begin{align}
    \op{\Lambda}^+\opSP[\vec n]\op{\Lambda}^+ & = 
    \op{\Lambda}^+\opSCz[\vec n]\op{\Lambda}^+ \,, \\
    \op{\Lambda}^-\opSP[\vec n]\op{\Lambda}^- & = 
    \op{\Lambda}^-\opSCz[\vec n]\op{\Lambda}^- 
  \end{align}
\end{subequations}
hold.  Consequently, the Pauli and the Czachor spin operators yield
the same expectation values when applied to the subspaces of the
eigenstates \eqref{eq:pos_neg_free} of the free-particle Dirac
Hamiltonian $\opHo$ with positive energy or negative energy,
respectively.

We also note that the eigenfunctions of the Foldy-Wouthuysen spin
operator with momentum strictly perpendicular or strictly parallel to
the spin orientation $\vec n$ are eigenfunctions of the Czachor spin
operator, too.  Furthermore, the total squared length of the spin
operator $\opSCz$ is $\opSCz^2 = (3m_0^2c^2 +
\opp^2)/(4m_0^2c^2+4\opp^2)$.  The fact that it is shorter than the
squared length of the Pauli operator $\opSP^2=3/4$ was associated with
the Lorentz contraction in \cite{Czachor:1997:relativistic_EPR}.  In
particular, for an ultrarelativistic particle, that is, $|\vec
p|\to\infty$, we have $\opSCz^2\to1/4$.  In this limit, the spin
components in the two directions perpendicular to $\vec p$ vanish.

\subsection{Frenkel spin operator}

A fourth definition of the spin is the quantum mechanical analog of a
classical spin vector as studied originally by Frenkel
\cite{Pryce:1948:Mass-Centre,Wightman:1960,Hilgevoord:Wouthuysen:1963:Dirac_spin,Frenkel:1926:Elektrodynamik_des,Bargmann:etal:1959:Precession}
\begin{equation}
  \label{eq:SF}
  \opSF = \frac{1}{2}\opSigma +
  \frac{\I\beta}{2m_0c}\opp\times\vec\alpha \,.
\end{equation}
It also commutes with the free Dirac Hamiltonian
\begin{equation}
  \commut{\opHo}{\opSF[\vec n]}=0 \,,
\end{equation}
but similarly to the Czachor operator
it does not obey the angular momentum algebra, viz.,
\begin{equation}
  \commut{\opSF[i]}{\opSF[j]} = 
  \I\varepsilon_{i,j,k}\left(
    \opSF[k] + \frac{\opSigma\cdot\op{\vec p}}{2m_0^2c^2}\op{p}_k
  \right)\,.
\end{equation}
The degenerate eigenvalues $s_\mathrm{F}$ and the normalized orthogonal
eigenvectors $\vec s_\mathrm{F}$ of $\opSF[\vec n]$ are given
by
\begin{align}
  \nonumber
  s_{\mathrm{F}\uparrow} & = \frac{p_0(\vec p_\perp)}{2m_0c} \,: & 
  \vec s_{\mathrm{F}\uparrow,1} & =
  \vec u_{\vec\chi_\uparrow, \vec p_\perp}\e^{\I\vec p\cdot \vec r} \,, &
  \vec s_{\mathrm{F}\uparrow,2} & =
  \vec v_{\vec\chi_\uparrow, \vec p_\perp}\e^{\I\vec p\cdot \vec r} \,, \\
  s_{\mathrm{F}\downarrow} & = -\frac{p_0(\vec p_\perp)}{2m_0c} \,: & 
  \vec s_{\mathrm{F}\downarrow,1} & =
  \vec u_{\vec\chi_\downarrow, \vec p_\perp}\e^{\I\vec p\cdot \vec r} \,, &
  \vec s_{\mathrm{F}\downarrow,2} & =
  \vec v_{\vec\chi_\downarrow, \vec p_\perp}\e^{\I\vec p\cdot \vec r} \,,
\end{align}
where $\vec p_\perp$ is the component of the momentum vector $\vec p$
that is perpendicular to $\vec n$.  Because the Frenkel spin operator
and the Pauli spin operator have different eigenvalues, the Frenkel
spin operator cannot be related to the Pauli spin operator via a
similarity transform.  This means that there is no operator $\op
T_{F}$ such that $\opSF$ equals $\op T_\mathrm{F}\opSP\op
T_\mathrm{F}^{-1}$.  We also note that the total squared length of the
Frenkel spin operator is $\opSF^2
=\mbox{$(3m_0^2c^2+2\opp^2)/(4m_0^2c^2)$}$.  The magnitude of the
eigenvalues and the total squared length $\opSF^2$ increase to infinity
as the momentum grows.

\subsection{Chakrabarti spin operator}

A fifth proposal for the spin operator has been introduced by
Chakrabarti 
\cite{Chakrabarti:1963:Canonical_Form,Gursey:1965,Gursey:1965:Equivalent_formulations,Choi:2013:Relativistic_spin_operator}.
It is defined via the similarity transformation
\begin{equation}
  \opSCh=\op{T}_\mathrm{Ch}\opSP\op{T}_\mathrm{Ch}^{-1} 
\end{equation}
that is induced by the antiunitary Lorentz boost operator
\begin{subequations}
  \begin{align}
    \op{T}_\mathrm{Ch} & =
    \frac{\oppo+m_0c+\vec{\alpha}\cdot\opp}{[2m_0c(\oppo+m_0c)]^{1/2}} \\
    \intertext{and its inverse, which is explicitly given by}
    \op{T}_\mathrm{Ch}^{-1} & =
    \frac{\oppo+m_0c-\vec{\alpha}\cdot\opp}{[2m_0c(\oppo+m_0c)]^{1/2}}\,.
  \end{align}
\end{subequations}
The explicit form of the (non-Hermitian) Chakrabarti spin operator
follows as
\begin{equation}
  \label{eq:SCh}
  \opSCh = 
  \frac{1}{2}\opSigma + 
  \frac{\I}{2m_0c}\vec\alpha\times\opp +
  \frac{1}{2m_0c(m_0c+\oppo)}
  \opp\times(\opSigma\times\opp)\,.
\end{equation}
In
\cite{Ryder:1999:Relativistic_Spin,Kirsch:Ryder:Hehl:2001:Gordon_decompositions}
the so-called G{\"u}rsey-Ryder operator
\begin{equation}
  \label{eq:SCh2}
  \opSCh = \frac{\oppo}{2m_0c}\opSigma - 
  \frac{\opp\cdot\opSigma}{2m_0c(m_0c+\oppo)}\opp - 
  \frac{\I}{2m_0c}\opp\times\vec\alpha
\end{equation}
was considered, which is just another algebraic expression for the
Chakrabarti spin operator.

The antiunitary similarity transformation operator
$\op{T}_\mathrm{Ch}$ is also Hermitian and $\beta$-pseudo-unitary,
that is,
$\op{T}_\mathrm{Ch}^\dagger=\beta\op{T}_\mathrm{Ch}^{-1}\beta^{-1}$,
which may be simplified to
$\op{T}_\mathrm{Ch}=\beta\op{T}_\mathrm{Ch}^{-1}\beta$. The operator
$\op{T}_\mathrm{Ch}$ transforms the operator
$\beta(\oppo+\vec\alpha\cdot\opp)$ into a diagonal
momentum-independent form, viz.,
\begin{equation}
  \op{T}_\mathrm{Ch}\beta(\oppo+\vec\alpha\cdot\opp)\op{T}_\mathrm{Ch}^{-1}
  =m_0c\beta\,.
\end{equation}
Similarly, when applied to the free Dirac Hamiltonian $\opHo$ the
operator $\op{T}_\mathrm{Ch}$ makes it almost diagonal, viz.,
\begin{equation}
  \label{eq:H0_Ch}
  \op{T}_\mathrm{Ch}\opHo\op{T}^{-1}_\mathrm{Ch}=c\beta\oppo +\op{h}\,,
\end{equation}
where $\op{h}$ is in the Dirac representation the matrix
\begin{equation}
  \label{eq:zeta}
  \op{h} =
  2 c\begin{pmatrix}
    0 & 0 \\
    \vec{\sigma}\cdot\opp & 0
  \end{pmatrix}\,.
\end{equation}
Here $0$ denotes a $2\times2$ zero matrix.  Note that the transformed
Hamiltonian \eqref{eq:H0_Ch} is not Hermitian as a consequence of
$\op{T}_\mathrm{Ch}$ not being unitary.

We also note that the operator $\beta\opSCh$ is Hermitian with respect
the usual scalar product, thus $\opSCh$ is $\beta$-pseudo-Hermitian.
Because $\opSCh$ originates from a similarity transformation of $\opSP$
it satisfies
\begin{equation}
  \commut{\opSCh[i]}{\opSCh[j]}=\I\varepsilon_{i,j,k}\opSCh[k] \,;
\end{equation}
its squared length is $\opSCh^2=3/4$, but its time evolution is
nontrivial because of the nonvanishing commutator
\begin{equation}
  \commut{\opHo}{\opSCh[\vec n]} =
  \vec n\cdot \left(
    \I \vec\alpha\times\opp \left(
      \frac{\oppo}{m_0} + c\beta
    \right) +
    \frac{\opp\times(\opSigma\times\opp)}{m_0} 
  \right) \,.
\end{equation}
The degenerate eigenvalues
$s_\mathrm{Ch}$ and the normalized eigenvectors $\vec s_\mathrm{Ch}$ of
$\opSCz[\vec n]$ follow directly via $\vec
s_{\mathrm{Ch}}=\op{T}_\mathrm{Ch}\vec s_{\mathrm{P}}$ (see
(\ref{eq:Pauli_s})) and are given by
\begin{align}
  \nonumber
  s_{\mathrm{Ch}\uparrow} & = \frac 12 \,: & 
  \vec s_{\mathrm{Ch}\uparrow,1} & =
  \vec u_{\vec\chi_\uparrow, \vec p}\e^{\I\vec p\cdot \vec r} \,, &
  \vec s_{\mathrm{Ch}\uparrow,2} & =
  \vec v_{\vec\chi_\uparrow, -\vec p}\e^{\I\vec p\cdot \vec r} \,, \\
  s_{\mathrm{Ch}\downarrow} & = -\frac 12 \,: & 
  \vec s_{\mathrm{Ch}\downarrow,1} & =
  \vec u_{\vec\chi_\downarrow, \vec p}\e^{\I\vec p\cdot \vec r} \,, &
  \vec s_{\mathrm{Ch}\downarrow,2} & =
  \vec v_{\vec\chi_\downarrow, -\vec p}\e^{\I\vec p\cdot \vec r} \,.
\end{align}
Note that the Chakrabarti spin operator shares two of its eigenvectors
with the Foldy-Wouthuysen spin operator, viz., $\vec
s_{\mathrm{Ch}\uparrow,1}=\vec s_{\mathrm{FW}\uparrow,1}$ and $\vec
s_{\mathrm{Ch}\downarrow,1}=\vec s_{\mathrm{FW}\downarrow,1}$.
Therefore, the Foldy-Wouthuysen spin operator and the Chakrabarti spin
operator are equivalent when applied to the subspace of the eigenstates
(\ref{eq:pos_free}) of the free-particle Dirac Hamiltonian with positive
energy.  In other words, the operator identity
\begin{equation}
  \label{eq:SFW_SCh_op}
  \opSFW[\vec n]\op{\Lambda}^+=\opSCh[\vec n]\op{\Lambda}^+
\end{equation}
holds, which also follows by comparing the expressions
(\ref{eq:SFW_bar_bar_bar}) and (\ref{eq:SCh2}) for the Foldy-Wouthuysen
and the spin operators.  In contrast, the operators $\opSFW[\vec
n]\op{\Lambda}^-$ and $\opSCh[\vec n]\op{\Lambda}^-$ are \emph{not}
equivalent, however, the following operator equality holds
\begin{equation}
  \label{eq:SFW_SCh-}
  \op{\Lambda}^-\opSFW[\vec n]\op{\Lambda}^- = 
  \op{\Lambda}^-\opSCh[\vec n]\op{\Lambda}^- \,.
\end{equation}

As a consequence of the non-Hermiticity, the eigenvectors of the
Chakrabarti spin operator are not all pairwise orthogonal to each
other, in particular not the states $ \vec s_{\mathrm{Ch}\uparrow,2}$
and $\vec s_{\mathrm{Ch}\downarrow,2}$, which have \emph{different}
spin orientations.  Furthermore, expectation values of the Chakrabarti
spin operator can lie outside the spectral range between $-1/2$ and
$1/2$.  For example, for the Gaussian wave packet of momentum width
$\sigma$ and mean momentum $\bar{\vec p}=\trans{(\bar p_x, 0, 0)}$,
\begin{equation}
  \Psi(\vec r) = 
  \int \frac{1}{\Big((2\pi)\sqrt{2\pi}\sigma\Big)^{3/2}}
  \exp\left(\I\vec p\cdot\vec r -\frac{(\vec p-\bar{\vec p})^2}{4\sigma^2}\right)
  \begin{pmatrix}
    1 \\ 0 \\ 0 \\ 0
  \end{pmatrix}\,\D^3 p
\end{equation}
we obtain for the expectation value of the Chakrabarti spin in the
$z$~direction
\begin{multline}
  \braket{\Psi| \opSCh[3]| \Psi} =\\
  \frac{1}{2} + 
  \frac{1}{\sqrt{2\pi}\sigma}\int
  \exp\left( -\frac{(p_x-\bar p_x)^2}{2\sigma^2}\right)
  \frac{p_x^2/(m_0c)^2}{2\sqrt{p_x^2/(m_0c)^2 + 1}}
  \,\D p_x\,.
\end{multline}
In the limit $\sigma\to0$ we find 
\begin{equation}
  \braket{\Psi| \opSCh[3]| \Psi} =
  \frac{1}{2} + 
  \frac{\bar p_x^2/(m_0c)^2}{2\sqrt{\bar p_x^2/(m_0c)^2 + 1}}\,,
\end{equation}
which is strictly larger than $1/2$ and grows in leading order linearly
as $|\bar p_x|\to\infty$.  This example also illustrates that the
expectation values $\braket{\Psi|\opSFW[\vec n]|\Psi}$ and
$\braket{\Psi|\opSCh[\vec n]|\Psi}$ are \emph{not} equal for general
states, as it was claimed recently in
\cite{Choi:2013:Relativistic_spin_operator}.

\subsection{Pryce spin operator}

A sixth proposal for a relativistic spin operator goes back to Pryce
\cite{Pryce:1935:Commuting_Co-Ordinates,Pryce:1948:Mass-Centre,Macfarlane:1963:Relativistic_Two_Particle,Ryder:1999:Relativistic_Spin},
who introduced the operator
\begin{equation}
  \label{eq:SPr_bar_bar}
   \opSPr = \frac{1}{m_0c}\left(
     \opW - \frac{\op{W}_0}{\opHo/c+m_0c}\opp
   \right)\,,
\end{equation}
which was applied in the context of quantum field theory
\cite{Bogolubov:1990:general} and relativistic quantum information
\cite{Terno:2003:relativistic_spin}.  Utilizing the definitions
\eqref{eq:Pauli-Lubanski_W} and \eqref{eq:Pauli-Lubanski_W0} the Pryce
spin operator's form is given by
\begin{equation}
  \label{eq:SPr}
  \opSPr = 
  \frac{1}{2}\beta\opSigma + 
  \frac{c\gamma^5(\beta+1)}{2(\opHo+m_0c^2)} \opp\,,
\end{equation}
where the $4\times4$ matrix $\gamma^5=\I\alpha_3\alpha_2\alpha_1$ is in
the Dirac representation defined as
\begin{equation}
  \gamma^5 =
  \begin{pmatrix}
    0 & \mathbb{I}_2 \\ \mathbb{I}_2 & 0 \\
  \end{pmatrix}\,.
\end{equation}
The equivalent expression for the Pryce spin operator
\begin{equation}
  \label{eq:SPr_bar}
  \opSPr = 
  \frac{1}{2\opHo}\left(
    m_0c^2\opSigma - \I c\beta\vec\alpha\times\opp +
    \frac{c^2\opp\cdot\opSigma}{\opHo+m_0c^2}\opp
  \right)
\end{equation}
was given in \cite{Pryce:1948:Mass-Centre}, which is almost identical to
the definition \eqref{eq:SFW_bar} of the Foldy-Wouthuysen spin operator
except that $c\oppo$ has been replaced by $\opHo$. Using the operator
identities
\begin{equation}
  \opHo^{-1} = 
  \frac{\vec{\alpha}\cdot\opp + m_0c\beta}{c\opp^2} 
\end{equation}
and
\begin{equation}
  (\opHo+m_0c^2)^{-1} = 
  \frac{\vec{\alpha}\cdot\opp + m_0c(\beta-1)}{c\opp^2} \,,
\end{equation}
we may turn (\ref{eq:SPr}) and (\ref{eq:SPr_bar}) into a form that is
more convenient for actual calculations, yielding
\begin{equation}
  \label{eq:SPr2}
  \opSPr = 
  \frac{1}{2}\beta\opSigma + 
  \frac{\gamma^5(\beta+1)\vec\alpha\cdot\opp}
  {2\opp^2}\opp
\end{equation}
or equivalently 
\begin{equation}
  \label{eq:SPr3}
  \opSPr = 
  \frac{1}{2}\beta\opSigma + 
  \frac{1}{2}\opSigma\cdot\opp(1-\beta)\dfrac{\opp}{\opp^2}\,.
\end{equation}
This form is unexpectedly simple and was suggested independently from
Pyrce by Stech in \cite{Stech:1956:Steuung}.  This form suggests that
the relativistic Pryce spin operator is a function of the momentum
operator's direction only, not depending on the mass $m_0$, the speed of
light $c$, or the amount of the momentum.  In the Dirac representation,
the operator is block diagonal, viz.,
\begin{equation}
  \opSPr = \frac{1}{2}\beta\opSigma + 
  \frac{1}{\opp^2}
  \begin{pmatrix}
    0 & 0 \\
    0 & \vec\sigma\cdot\opp
  \end{pmatrix}
  \opp\,.
\end{equation}
The spin operator $\opSPr$ fulfills the angular momentum algebra
\begin{equation}
  \commut{\opSPr[i]}{\opSPr[j]}=\I\varepsilon_{i,j,k}\opSPr[k] \,,
\end{equation}
its squared length is $\opSPr^2=3/4 $, and it commutes with the free Dirac
Hamiltonian $\commut{\opHo}{\opSPr}=0$. The degenerate eigenvalues
$s_\mathrm{Pr}$ and the orthogonal normalized eigenvectors of
$\opSPr[\vec n]$ are given by
\begin{align}
  \label{eq:SPr_eigen}
  \nonumber
  s_{\mathrm{Pr}\uparrow} & = \frac 12 \,: & 
  \vec s'_{\mathrm{Pr}\uparrow,1} & =
  \begin{pmatrix}
    \vec\chi_\uparrow \\ 0 
  \end{pmatrix}\e^{\I\vec p\cdot \vec r} \,, &
  \vec s'_{\mathrm{Pr}\uparrow,2} & =
  \begin{pmatrix}
    0 \\[1ex] \dfrac{\vec\sigma\cdot\vec p}{|\vec p|}\vec\chi_\uparrow
  \end{pmatrix}\e^{\I\vec p\cdot \vec r} \,, \\
  s_{\mathrm{Pr}\downarrow} & = -\frac 12 \,: & 
  \vec s'_{\mathrm{Pr}\downarrow,1} & =
    \begin{pmatrix}
    \vec\chi_\downarrow \\ 0 
  \end{pmatrix}\e^{\I\vec p\cdot \vec r} \,, &
  \vec s'_{\mathrm{Pr}\downarrow,2} & =
  \begin{pmatrix}
    0 \\[1ex] \dfrac{\vec\sigma\cdot\vec p}{|\vec p|}\vec\chi_\downarrow
  \end{pmatrix}\e^{\I\vec p\cdot \vec r} \,.
\end{align}
The eigenvectors in (\ref{eq:SPr_eigen}) have a particular simple
form; however, they are not simultaneously eigenvectors of the free
Dirac Hamiltonian $\opHo$ and $\opSPr[\vec n]$. The simultaneous
eigenvectors of both operators can be written in the matrix form
\begin{multline}
  \begin{pmatrix}
    \vec s_{\mathrm{Pr}\uparrow,1} &
    \vec s_{\mathrm{Pr}\downarrow,1} &
    \vec s_{\mathrm{Pr}\uparrow,2} &
    \vec s_{\mathrm{Pr}\downarrow,2} 
  \end{pmatrix}= \\
  \begin{pmatrix}
    \vec u_{\vec\upsilon_\uparrow, \vec p} &
    \vec u_{\vec\upsilon_\downarrow, \vec p} &
    \vec v_{\vec\upsilon_\uparrow, \vec p} &
    \vec v_{\vec\upsilon_\downarrow, \vec p} 
  \end{pmatrix}
  \begin{pmatrix}
    \begin{pmatrix}
      \vec\chi_\uparrow & \vec\chi_\downarrow 
    \end{pmatrix} &
    \begin{pmatrix}
      0 & 0 \\ 0 & 0
    \end{pmatrix} \\
    \begin{pmatrix}
      0 & 0 \\ 0 & 0
    \end{pmatrix} &
     \dfrac{\vec\sigma\cdot\vec p}{|\vec p|} 
     \begin{pmatrix}
       \vec\chi_\uparrow & \vec\chi_\downarrow 
     \end{pmatrix}
  \end{pmatrix}\e^{\I\vec p\cdot \vec r}
\end{multline}
with $\vec\upsilon_\uparrow=\trans{(1,0)}$ and
$\vec\upsilon_\downarrow=\trans{(0,1)}$.  The eigenvectors $\vec
s_{\mathrm{Pr}\uparrow,1}$ and $\vec s_{\mathrm{Pr}\downarrow,1}$ have
the positive energy-eigenvalue $cp_0(\vec p)$, whereas $\vec
s_{\mathrm{Pr}\uparrow,2}$ and $\vec s_{\mathrm{Pr}\downarrow,2}$ have
the negative energy-eigenvalue $-cp_0(\vec p)$.  Note that $\vec
s'_{\mathrm{Pr}\uparrow,1}$ and $\vec s'_{\mathrm{Pr}\downarrow,1}$
are also eigenvectors of the Pauli spin operator, while $\vec
s_{\mathrm{Pr}\uparrow,1}$ and $\vec s_{\mathrm{Pr}\downarrow,1}$ are
also eigenvectors of the Foldy-Wouthuysen spin operator.

Because $\vec s_{\mathrm{Pr}\uparrow,1}$ and $\vec
s_{\mathrm{Pr}\downarrow,1}$ are positive-energy eigenfunctions of the
free Dirac Hamiltonian too, the operator identity
\begin{equation}
  \label{eq:pos_SFW_SPr}
  \opSFW[\vec n]\op{\Lambda}^+ = 
  \opSPr[\vec n]\op{\Lambda}^+
\end{equation}
holds. Note, however, that the operators $\op{\Lambda}^-\opSFW[\vec
n]\op{\Lambda}^-$ and $\op{\Lambda}^-\opSPr[\vec n]\op{\Lambda}^-$ are
\emph{not} equivalent but states from different energy subspaces are
coupled identically, viz.,
\begin{equation}
  \Lambda^+ \opSFW\Lambda^- = \Lambda^+ \opSPr\Lambda^- =
  \Lambda^- \opSFW\Lambda^+ = \Lambda^- \opSPr\Lambda^+ \,.
\end{equation}
As $\opSP[\vec n]$ and $\opSPr[\vec n]$ share the same eigenvalues, it
is possible to express the Pryce operator as a similarity transformation
based on $\opSP$, similar to the corresponding transformations for
$\opSFW$ and $\opSCh$.  In fact, one can show that $\opSPr=\op
T_{\mathrm{Pr}}\opSP \op T_{\mathrm{Pr}}^{-1}$ with the unitary operator
\begin{equation}
  \op T_{\mathrm{Pr}} =
  \begin{pmatrix}
    \mathbb{I}_2 & 0 \\[1ex]
    0 & \I\dfrac{\vec\sigma\cdot\opp}{|\opp|}
  \end{pmatrix}\,.
\end{equation}
Utilizing this transformation, it is possible to formulate relativistic
quantum mechanics in a representation where the free Hamiltonian takes
the form
\begin{equation}
  \op{T}_\mathrm{Pr}^{-1}\opHo\op{T}_\mathrm{Pr} = 
  \left(mc^2 - \I\gamma^5 c |\opp|\right)\beta
\end{equation}
similar to the Foldy-Wouthuysen picture.

\subsection{Fradkin-Good operator}

An operator that has a definition similar to the Pryce operator was
considered by Fradkin and Good in
\cite{Fradkin:Good:1961:Polarization_Operators}, the so-called
polarization vector, which is defined as
\begin{equation}
  \label{eq:Spol}
  \opSFG = 
  \frac{1}{2}\beta\opSigma + 
  \frac{1}{2}\opSigma\cdot\opp\left(\frac{\opHo}{c\oppo}-\beta\right)\dfrac{\opp}{\opp^2}\,.
\end{equation}
It has been studied extensively in
\cite{Fradkin:Good:1961:Polarization_Operators,Kirsch:Ryder:Hehl:2001:Gordon_decompositions}.
The squared length of $\opSFG$ is $\opSFG^2=3/4 $ and it commutes with
the free Dirac Hamiltonian $\commut{\opHo}{\opSPr}=0$, but it does not
fulfill the angular momentum algebra; rather
\begin{equation}
  \commut{\opSFG[i]}{\opSFG[j]} = 
  \I\varepsilon_{i,j,k}\opSFG[k] \frac{\opHo}{c\oppo}\,.
\end{equation}
The degenerate eigenvalues $s_\mathrm{FG}$ and the normalized orthogonal
eigenvectors $\vec s_\mathrm{FG}$ of $\opSFG[\vec n]$ and $\opHo$ are
given by
\begin{align}
  \label{eq:SFG_eigen}
  \nonumber
  s_{\mathrm{FG}\uparrow} & = \frac 12 \,: & 
  \vec s_{\mathrm{FG}\uparrow,1} & =
  \vec u_{\vec\chi_\uparrow, \vec p}\e^{\I\vec p\cdot \vec r} \,, &
  \vec s_{\mathrm{FG}\uparrow,2} & =
  \vec v_{\vec\chi_\downarrow, \vec p}\e^{\I\vec p\cdot \vec r}\,, \\
  s_{\mathrm{FG}\downarrow} & = -\frac 12 \,: & 
  \vec s_{\mathrm{FG}\downarrow,1} & =
  \vec u_{\vec\chi_\downarrow, \vec p}\e^{\I\vec p\cdot \vec r} \,, &
  \vec s_{\mathrm{FG}\downarrow,2} & =
  \vec v_{\vec\chi_\uparrow, \vec p}\e^{\I\vec p\cdot \vec r} \,.
\end{align}
The eigenfunctions of $\opSFG[\vec n]$ are also eigenfunctions of
$\opSFW[\vec n]$ but in the case of negative-energy states with opposite
eigenvalues.  Thus
\begin{align}
  \opSFW[\vec n]\op{\Lambda}^+ & = 
  \opSFG[\vec n]\op{\Lambda}^+ \,, \\
  \opSFW[\vec n]\op{\Lambda}^- & = -
  \opSFG[\vec n]\op{\Lambda}^- \,.
\end{align}
Comparing the definitions (\ref{eq:SPr3}) and (\ref{eq:Spol})
immediately reveals the operator identity 
\begin{equation}
  \opSPr[\vec n]\op{\Lambda}^+ = 
  \opSFG[\vec n]\op{\Lambda}^+ \,.
\end{equation}
The Fradkin-Good operator $\opSFG$ is not related to the
Pauli operator $\opSP$ via a similarity transformation; instead 
\begin{equation}
  \opSFG =
  \op{T}_\mathrm{FW}\beta\opSP\op{T}_\mathrm{FW}^{-1} 
\end{equation}
with $\op{T}_\mathrm{FW}$ as defined in \eqref{eq:UFW}.

\subsection{Applying spin operators to wave functions}

The spin operators in Table~\ref{tab:properties} are defined as
functions of the canonical momentum operator $\op{\vec p}$, which has
in position space representation the form $\op{\vec p}=-\I\vec\nabla$.
Consequently, the spin operators in Table~\ref{tab:properties} are
rather complicated differential operators and their application to
position space wave functions $\Psi(\vec r)$ is not straightforward.
Furthermore, the spin operators are not gauge independent as
expectation vales of the canonical momentum operator depend on the
choice of the gauge.  One can deal with both issues in the following
way.

Noting that in canonical momentum space the canonical momentum operator
$\op{\vec p}$ is just a real-valued vector and that none of the proposed
spin operators depends on position, one can apply the spin operators
easily to wave functions given in canonical momentum space
\begin{equation}
  \tilde\Psi(\vec p, t) = 
  \frac{1}{(2\pi)^{3/2}} \int 
  \Psi(\vec r, t) \exp(-\I\,\vec r\cdot\vec p)\,\D^3r \,,
\end{equation}
where the operators in Table~\ref{tab:properties} become plain
matrices.  The spin operators as proposed in
Table~\ref{tab:properties} can represent operators that correspond to
measurable observables only in gauges with $\vec A(\vec r, t)=0$ where
the canonical momentum operator equals the physical kinematic momentum
operator.  Physical spin operators for general gauges with $\vec
A(\vec r, t)\ne0$, however, can be obtained by replacing $\op{\vec p}$
by $\op{\vec p}-q\vec A(\vec r, t)$ in the definitions in
Table~\ref{tab:properties}.  In this way the spin operators become
position dependent and consequently the canonical momentum
representation (Fourier representation) of the spin operators are no
longer plain matrices and therefore difficult to apply whether the
wave function is given in position space or Fourier space.  For this
reason we will concentrate below on physical setups with vanishing
vector potential.

\section{Time dependence of the spin in scattering dynamics}
\label{sec:scattering}

Summarizing the results of Sec.~\ref{sec:spins}, the following
operator identities hold:
\begin{gather}
  \Lambda^+\opSP\Lambda^+ = \Lambda^+ \opSCz\Lambda^+ \,,\\
  \opSFW\Lambda^+ = \opSCh\Lambda^+  = \opSPr\Lambda^+ = \opSFG\Lambda^+ \,.
\end{gather}
Thus, the Pauli and the Czachor spin operators yield the same
expectation values in the subspace free-particle states with positive
energy and also the Foldy-Wouthuysen, the Chakrabarti, the Pryce, and
the Fradkin-Good spin operators are equivalent in the positive-energy
subspace of free-particle states. If an interaction with some external
fields is introduced a superposition of positive-energy free-particle
solutions may evolve such that negative-energy free-particle states
become populated and therefore it is possible to distinguish between
the various spin operators by determining their expectation
values. For a detailed discussion of the quantum field-theoretic
interpretation of transitions to negative-energy states see
\cite{Krekora_Su_Grobe:2004:transitions_to_negative_Dirac_continuum}.

\subsection{Reflection of a wave packet  at a step potential}
\label{sec:step_potential}

\begin{figure}
  \centering
  \includegraphics[scale=0.55]{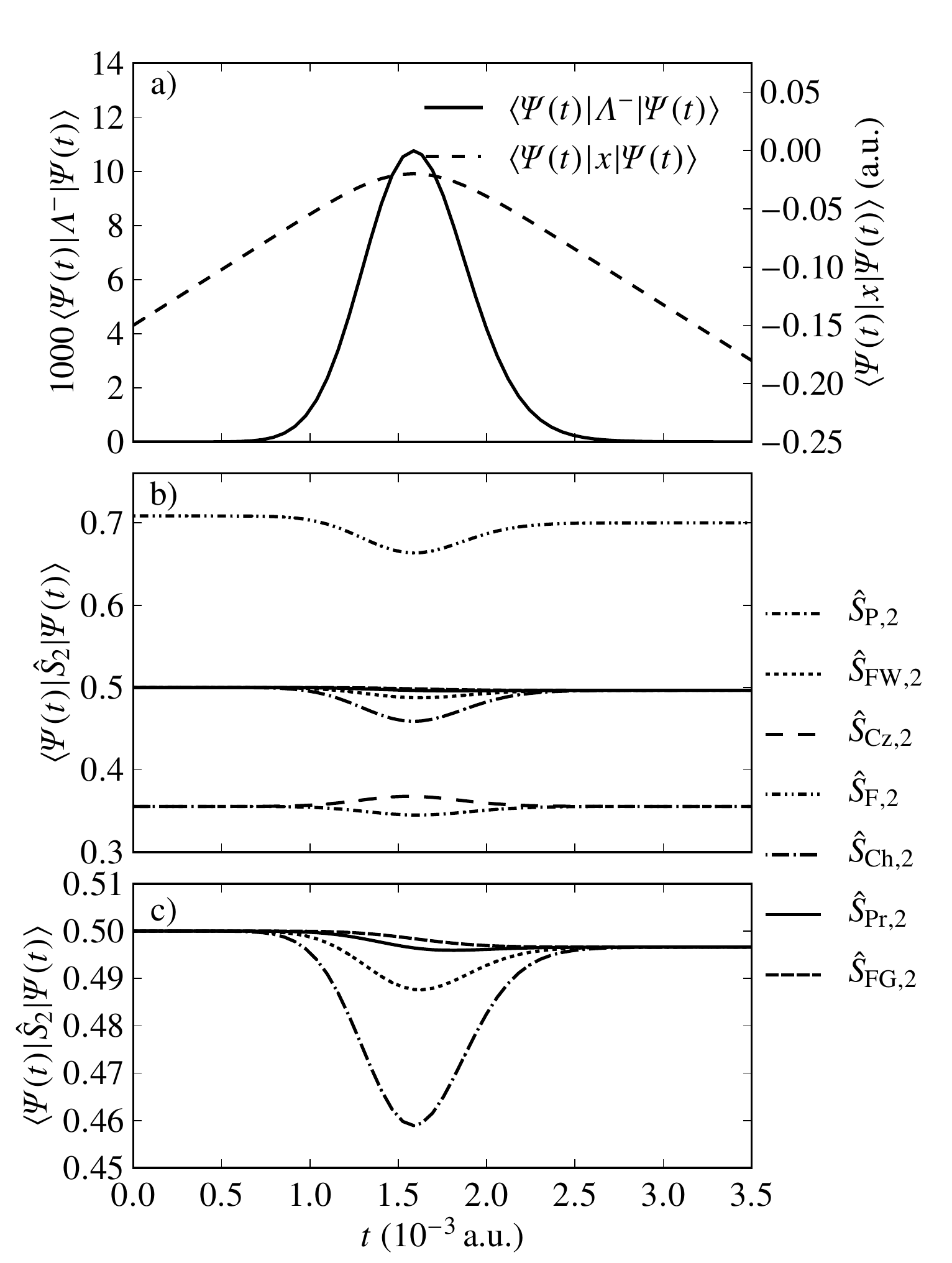}
  \caption{Relativistic spin dynamics in scattering at a step potential
    \eqref{eq:step}. Sub-figure~(a) shows the occupation probability
    $\braket{\Psi|\Lambda^-|\Psi}$ of negative-energy free-particles and
    the $x$-coordinate of the center of mass $\braket{\Psi|x|\Psi}$.
    Sub-figure~(b) presents the expectation values of various spin
    operators as a function of time $t$ and sub-figure~(c) shows a
    magnification of sub-figure~(b). All results given in atomic units,
    particle's mass equals $m_0=1\,\mathrm{a.u.}$}
  \label{fig:step_potential}
\end{figure}

As a first example let us consider the relativistic spin dynamics in
scattering at a smooth two-dimensional step potential
\begin{equation}
  \label{eq:step}
  q\phi(x,y) = \frac{V_0}{2}\left(
    1+\tanh \frac xw
  \right)
\end{equation}
with $V_0=1.95m_0c^2$ and $w=1/(4c)$ (in atomic units) such that the
barrier is high but still below the critical value $2m_0c^2$ that
would permit Klein tunneling \cite{Klein:1929:Reflexion}.  The initial
state is a Gaussian superposition of common eigenstates of $\opp$,
$\opHo$, and $\opSFW[2]$ having positive energy and positive spin in
the $y$~direction
\begin{equation}
  \label{eq:step_init}
  \Psi(\vec r, 0) = 
  \frac{1}{2\pi}\int g(\vec p') \vec u_{\vec\chi_\uparrow,\vec p'} 
  \e^{\I\vec p'\cdot \vec r} \,\D^2 p'\,,
\end{equation}
with $\vec r=\trans{(x, y)}$, $\vec p'=\trans{(p'_x, p'_y)}$, and
$\vec\chi_\uparrow=\trans{(1,\I)}/\sqrt2$ here and $g(\vec p')$
denoting a Gaussian weight function corresponding to a spatial width
of $0.025$\,a.u.\ in the $x$~and $y$~directions.  The wave packet's
initial center of mass is at
$\trans{(-0.175\mathrm{\,a.u.},0\mathrm{\,a.u.})}$ and its initial
mean momentum is $\trans{(m_0c,0)}$ such that the two-dimensional wave
packet approaches the barrier from the left.  The quantum dynamics is
simulated by solving the time-dependent Dirac equation numerically by
a Fourier split operator method
\cite{Braun:etal:1999:Numerical_approach,Bauke:Keitel:2011:Accelerating}. When
the wave packet interacts with the barrier negative-energy states
become occupied as indicated by the quantity
$\braket{\Psi(t)|\Lambda^-|\Psi(t)}$ in
Fig.~\ref{fig:step_potential}(a). After reflection when the wave
packet has left the interaction zone, however, negative-energy states
are no longer occupied.

We determine spin expectation values in the $y$~direction.  All spin
expectation values change during the interaction with the potential
step.  Initially, the Pauli and the Czachor spin operators yield the
same expectation values as a result of the initial condition and the
Foldy-Wouthuysen, the Chakrabarti, the Pryce, and the Fradkin-Good
spin operators all give an expectation value of $1/2$, see
Fig.~\ref{fig:step_potential}(b).  When the wave packet interacts with
the step potential and a substantial fraction of free-particle
negative-energy states is occupied the spin expectation values change
in a specific way such that they differ for all proposed spin
operators.  In particular, the spin expectation values for the
Foldy-Wouthuysen, the Chakrabarti, the Pryce, and the Fradkin-Good
spin operators are reduced.  In the cases of the Pryce and the
Fradkin-Good spin operators the change of the expectation value is so
small (about 1\,\%) that it is not visible on the scale of
Fig.~\ref{fig:step_potential}(b) and therefore
Fig.~\ref{fig:step_potential}(c) shows a magnification of
Fig.~\ref{fig:step_potential}(b).

\begin{table}
  \caption{Initial and final spin expectation values for the scattering
    dynamics in Fig.~\ref{fig:step_potential}. In contrast to other
    operators, the Pauli and the Czachor spin operators predict that the
    spin expectation value after interaction with the barrier is the
    same as before.}
  \label{tab:step_potential}
  \begin{ruledtabular}
    \begin{tabular}{@{}lcc@{}}
      Operator & 
      $\braket{\psi_{\uparrow}|\hat{S}_{2}|\psi_{\uparrow}}$ at
      $t=0$\,a.u. & 
      $\braket{\psi_{\uparrow}|\hat{S}_{2}|\psi_{\uparrow}}$ at
      $t=0.0035$\,a.u. \\
      \colrule
      $\opSP$ & 0.3556 & 0.3556 \\
      $\opSFW$ & 0.5000 & 0.4966 \\
      $\opSCz$ & 0.3556 & 0.3556 \\
      $\opSF$ & 0.7084 & 0.7001 \\
      $\opSCh$ & 0.5000 & 0.4966 \\
      $\opSPr$ & 0.5000 & 0.4966 \\
      $\opSFG$ & 0.5000 & 0.4966 \\
    \end{tabular}
  \end{ruledtabular}
\end{table}

The net effect of the scattering dynamics on the expectation value of
the spin depends on the spin operator as shown in
Table~\ref{tab:step_potential}.  While the Foldy-Wouthuysen, the
Chakrabarti, the Pryce, and the Frenkel spin operators predict a small
change of the spin's expectation value as a net effect of the step
potential, the Pauli and the Czachor spin operators predict that the
spin expectation value after interaction with the barrier is the same as
before.

\subsection{Spin dynamics in standing laser fields}
\label{sec:standing_laser_fields}

A similar distinction among the spin operators can be observed in a
rather different scattering dynamics of an electron in a standing wave
formed by two monochromatic laser fields.  In contrast to the prior
example in Sec.~\ref{sec:step_potential}, the electron has sharp
momentum and is therefore spatially delocalized and also the
interaction region is infinitely extended. Such a dynamic can lead to
the so-called Kapitza-Dirac effect
\cite{Kapitza:Dirac:1933:effect,Ahrens:etal:2012:spin_dynamics,Ahrens:etal:2013:Kapitza-Dirac}.

The system is described by the Dirac equation \eqref{eq:Dirac}, where
the effect of the time-averaged laser field can be modeled by the
ponderomotive potential \cite{Batelaan:2000:Kapitza-Dirac}
\begin{equation}
  \label{eq:ponderomotive}
  q\phi(\vec r,t) = V_0\cos^2(\vec k \cdot \vec r)w(t)\,.
\end{equation}
Here $V_0$ is the potential amplitude, $\vec k$ is the laser's wave
vector, and $w(t)$ denotes the temporal envelope of the standing light
wave.  As a consequence of the infinite extension of the periodic
laser field, only discrete subsets of momenta are coupled.  This allows
us to expand the quantum wave function in a basis of Foldy-Wouthuysen
spin operator eigenfunctions
\begin{multline}
  \vec \psi(\vec r,t) = \sum_{n} \Big(
   c_n^{+\vec\chi_\uparrow}(t) \vec u_{\vec\chi_\uparrow, \vec p} +
   c_n^{+\vec\chi_\downarrow}(t) \vec u_{\vec\chi_\downarrow, \vec p} \\
   + c_n^{-\vec\chi_\uparrow}(t) \vec v_{\vec\chi_\uparrow, \vec p} +
   c_n^{-\vec\chi_\downarrow}(t) \vec v_{\vec\chi_\downarrow, \vec p}
   \Big) \e^{\I(\vec p + n \vec k)\cdot \vec r} \,.
\end{multline}
Inserting this ansatz into the time-dependent Dirac equation, we find
a coupled set of ordinary differential equations for the amplitudes of
each mode. These differential equations are solved numerically with
the initial condition that all amplitudes are zero except
$c_0^{+\vec\chi_\uparrow}(0) = 1$.  (See
\cite{Ahrens:etal:2013:Kapitza-Dirac} for technical details.)

\begin{figure}
  \centering
  \includegraphics[scale=0.55]{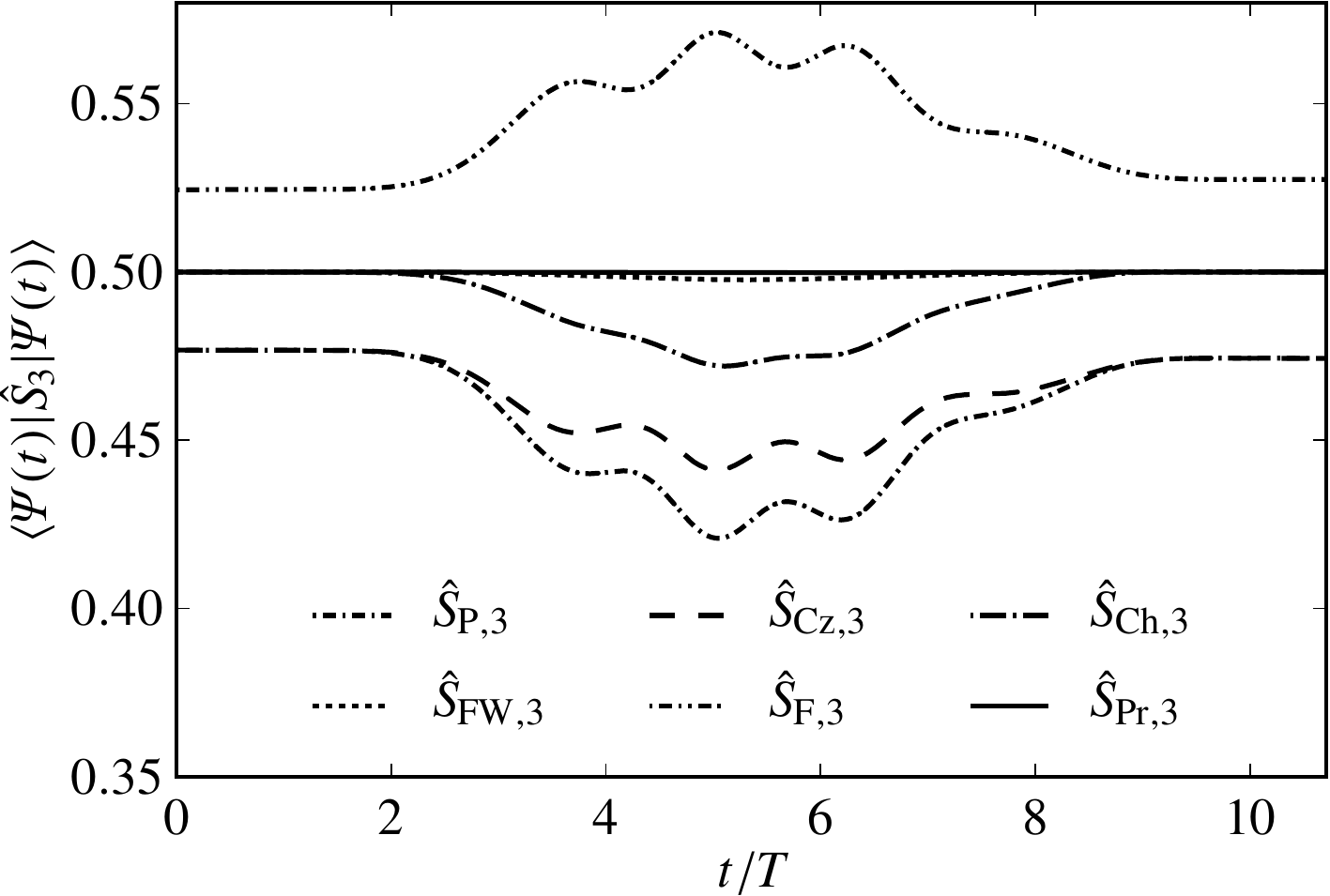}
  \caption{Spin expectation values for Kapitza-Dirac scattering in a
    standing laser field as a function of time (in units of the laser
    period $T$) for parameters given in the text.}
  \label{fig:KDE}
\end{figure}

In Fig.~\ref{fig:KDE} we give a specific realization of a spin dynamics
in a standing laser field, where the corresponding electric field vector
points in the $z$~direction, $V_0 = 0.88m_0 c^2$, and the wave vector
points in the $x$~direction, $\vec k = \trans{(0.5,0,0)} m_0 c$.  The
temporal envelope function was given by $w(t)=\sin^2(\pi
t/t_\mathrm{end})$ with $0<t<t_\mathrm{end}=10.7T$ and $T$ denoting the
laser period.  The electron's initial momentum is $\vec
p=\trans{(-0.3169,0,0.1)} m_0 c$ and the spin is initially oriented
in the $z$~direction, i.\,e., $\vec\chi_\uparrow = \trans{(1,0)}$.
When the full time dependence of the laser field is taken into account
(instead of just the ponderomotive model potential
\eqref{eq:ponderomotive}) these parameters lead to the relativistic
three-photon Kapitza-Dirac effect as investigated in
\cite{Ahrens:etal:2012:spin_dynamics}.

In close analogy to the dynamics in Sec.~\ref{sec:step_potential} we
find again three different values for the expectation value of the
initial spin for time $t=0$ (where the ponderomotive potential
\eqref{eq:ponderomotive} is zero) depending on the choice of the spin
operator (see Fig.~\ref{fig:KDE}).  The deviation from $1/2$ is
smaller than in Fig.~\ref{fig:step_potential} due to the less
relativistic electron momentum here.  For times where the
ponderomotive potential \eqref{eq:ponderomotive} is nonzero we observe
a time evolution of the spin expectation values qualitatively similar
to that in Fig.~\ref{fig:step_potential}.  During the interaction with
the laser field these values vary in an oscillatory fashion over time
and all differ significantly from each other except those associated
with the Foldy-Wouthuysen and the Pryce operators, which remain almost
constant over time.  Again the discrepancy in the expectation values
can be traced back to significant transient excitations of
negative-energy free-particle states as represented by the amplitudes
$c_n^{-\vec\chi_\uparrow}(t)$ and $c_n^{-\vec\chi_\downarrow}(t)$.

\section{Relativistic spin of the electron in the hydrogenic bound states}
\label{sec:hydrogen}

\subsection{Hydrogenic bound states} 

In this section we are going to investigate the spin of hydrogenic bound
states for a highly charged ion at rest with respect to different
definitions of the relativistic spin operator.  The degenerate bound
states of the Dirac Hamiltonian for the Coulomb potential $q\phi(\vec r,
t) = -{Z}/{|\vec r|}$ with atomic number $Z$,
\begin{equation}
  \opHC = \opHo - \frac{Z}{|\vec r|} 
\end{equation}
are commonly expressed as simultaneous eigenstates $\psi_{n,\kappa,j,m}$
of $\opHC$, $\opJ^2$, $\opJ[3]$, and the so-called spin-orbit operator
$\op{K}=\mbox{$\beta\{\opSigma\cdot[\vec r\times(-\I\vec\nabla)+1)]\}$}$
fulfilling the eigenequations
\cite{Bethe:Salpeter:2008:Atoms,Thaller:2005:vis_QM_II}
\begin{subequations}
  \begin{align}
    \opHC \psi_{n,\kappa,j,m} & = \mathcal{E}(n, \kappa)\psi_{n,\kappa,j,m}\,, &
    n & = 1,2,\dots\,,\\
    \op{K} \psi_{n,\kappa,j,m} & = \kappa\psi_{n,\kappa,j,m}\,, & 
    |\kappa| & = 1,2,\dots,n \,,\kappa\ne-n\,,\\
    \opJ^2 \psi_{n,\kappa,j,m} & = j(j+1)\psi_{n,\kappa,j,m}\,, &
    j & = |\kappa|-\tfrac{1}{2} \,,\\
    \opJ[3] \psi_{n,\kappa,j,m} & = m\psi_{n,\kappa,j,m} \,, &
    m & = -j,(j-1),\dots,j \,.
  \end{align}
\end{subequations}
The eigenenergies $\mathcal{E}(n, \kappa)$ are given with
$\alpha_\mathrm{el}$ denoting the fine-structure constant by
\begin{equation}
  \label{eq:energy}
  \mathcal{E}(n, \kappa) = 
  m_0c^2\left[
    1 + \left(
      \tfrac{\alpha_\mathrm{el}^2Z^2}
      {n-|\kappa|+\sqrt{\kappa^2-\alpha_\mathrm{el}^2Z^2}}
    \right)
  \right]^{-{1}/{2}}\,.
\end{equation}
Each eigenfunction $\psi_{n,\kappa,j,m}$ belongs to one of two
manifolds: For $\kappa =j+{1}/{2}\in \{1,2,\dots,n\}$,
\begin{subequations}
  \label{eq:hydrogen_eigen}
  \begin{equation}
    \label{eq:hydrogen_eigen_1}
    \psi_{n,\kappa,j,m}(r,\theta,\phi) =
    \begin{pmatrix}
      g_{n,\kappa,j}(r)\sqrt{\frac{j+m}{2j}}Y_{j-1/2,m-1/2}(\theta, \phi) \\[1.5ex]
      g_{n,\kappa,j}(r)\sqrt{\frac{j-m}{2j}}Y_{j-1/2,m+1/2}(\theta, \phi) \\[1.5ex]
      -f_{n,\kappa,j}(r)\I\sqrt{\frac{j-m+1}{2j+2}}Y_{j+1/2,m-1/2}(\theta, \phi) \\[1.5ex]
      f_{n,\kappa,j}(r)\I\sqrt{\frac{j+m+1/2}{2j+2}}Y_{j+1/2,m+1/2}(\theta, \phi)
    \end{pmatrix}
  \end{equation}
  and for $\kappa =-j-{1}/{2}\in \{-1,-2,\dots,-(n-1)\}$,
  \begin{equation}
    \label{eq:hydrogen_eigen_2}
    \psi_{n,\kappa,j,m}(r,\theta,\phi) =
    \begin{pmatrix}
      -g_{n,\kappa,j}(r)\sqrt{\frac{j-m+1}{2j+2}}Y_{j+1/2,m-1/2}(\theta, \phi) \\[1.5ex]
      g_{n,\kappa,j}(r)\sqrt{\frac{j+m+1}{2j+2}}Y_{j+1/2,m+1/2}(\theta, \phi) \\[1.5ex]
      f_{n,\kappa,j}(r)\I\sqrt{\frac{j+m}{2j}}Y_{j-1/2,m-1/2}(\theta, \phi) \\[1.5ex]
      f_{n,\kappa,j}(r)\I\sqrt{\frac{j-m}{2j}}Y_{j-1/2,m+1/2}(\theta, \phi)
    \end{pmatrix}\,.
  \end{equation}
\end{subequations}
The radial functions $g_{n,\kappa,j}(r)$ and $f_{n,\kappa,j}(r)$ can be
expressed in terms of confluent hypergeometric functions
\cite{Greiner:2000:RQM,Bjorken:Drell:1964:RQM} and
$Y_{l,m}(\theta,\phi)$ denote the complex-valued orthonormal spherical
harmonics as defined in \cite{Arfken:Weber:2005:Mathematical_Methods}.

Explicitly, the degenerate hydrogenic ground state is
\cite{Bjorken:Drell:1964:RQM}
\begin{subequations}
  \label{eq:hydrogen_real}
  \begin{align}
    \label{eq:hydrogen_real_1}
    \psi_{1, 1, \frac{1}{2}, \frac{1}{2}}(r,\theta,\phi) & = \mathcal{N}\psi(r)
    \begin{pmatrix}
      Y_{0,0}(\theta,\phi) \\
      0 \\
      \I\frac{1-\gamma}{Z\alpha_\mathrm{el}}\sqrt{\frac13}Y_{1,0}(\theta,\phi) \\[1.5ex]
      -\I\frac{1-\gamma}{Z\alpha_\mathrm{el}}\sqrt{\frac23}Y_{1,1}(\theta,\phi)
    \end{pmatrix}\\
  \label{eq:hydrogen_real_2}
    \psi_{1, 1, \frac{1}{2}, -\frac{1}{2}}(r,\theta,\phi) & = \mathcal{N}\psi(r)
    \begin{pmatrix}
      0 \\
      Y_{0,0}(\theta,\phi) \\
      \I\frac{1-\gamma}{Z\alpha_\mathrm{el}}\sqrt{\frac23}Y_{1,-1}(\theta,\phi) \\[1.5ex]
      -\I\frac{1-\gamma}{Z\alpha_\mathrm{el}}\sqrt{\frac13}Y_{1,0}(\theta,\phi)
    \end{pmatrix}
  \end{align}
\end{subequations}
with $\gamma=\sqrt{1-Z^2\alpha_\mathrm{el}^2}$, the radial function
\begin{equation}
  \psi(r) = 
  % \frac{(2m_0Z\alpha)^{3/2}}{\sqrt{4\pi}}
  % \sqrt{\frac{1+\gamma}{2\GAMMA(1+2\gamma)}}
  \frac{\e^{-m_0Zr} }{(2m_0Zr)^{1-\gamma}}\,,
\end{equation}
the normalizing factor 
\begin{equation}
  \mathcal{N}=
  (2m_0Z)^{3/2}
  \sqrt{\frac{1+\gamma}{2\GAMMA(1+2\gamma)}} \,,
\end{equation}
and the electron rest mass $m_0$.  In momentum space, the degenerate
bound states (\ref{eq:hydrogen_real}) may be expressed as
\cite{Rubinowicz:1948:Coulomb_Momentum}
\begin{subequations}
  \label{eq:hydrogen_momentum}
  \begin{align}
    \label{eq:hydrogen_momentum_1}
    \tilde\psi_{1, 1, \frac{1}{2}, \frac{1}{2}}(p,\theta',\phi') & = \mathcal{N}
    \begin{pmatrix}
      \mathcal{J}_0(m_0Z,\gamma, p)Y_{0,0}(\theta',\phi') \\
      0 \\
      \frac{1-\gamma}{Z\alpha}\mathcal{J}_1(m_0Z,\gamma, p)\sqrt{\frac13}Y_{1,0}(\theta',\phi') \\[1.5ex]
      -\frac{1-\gamma}{Z\alpha}\mathcal{J}_1(m_0Z,\gamma,
      p)\sqrt{\frac23}Y_{1,1}(\theta',\phi')
    \end{pmatrix}\\
    \label{eq:hydrogen_momentum_2}
    \tilde\psi_{1, 1, \frac{1}{2}, -\frac{1}{2}}(p,\theta',\phi') & = \mathcal{N}
    \begin{pmatrix}
      0 \\
      \mathcal{J}_0(m_0Z,\gamma, p)Y_{0,0}(\theta',\phi') \\
      \frac{1-\gamma}{Z\alpha}\mathcal{J}_1(m_0Z,\gamma, p)\sqrt{\frac23}Y_{1,-1}(\theta',\phi') \\[1.5ex]
      -\frac{1-\gamma}{Z\alpha}\mathcal{J}_1(m_0Z,\gamma,
      p)\sqrt{\frac13}Y_{1,0}(\theta',\phi')
    \end{pmatrix}
  \end{align}
\end{subequations}
by using the functions $\mathcal{J}_0(z,\gamma, p)$ and
$\mathcal{J}_1(z,\gamma, p)$ as defined in (\ref{eq:J0}) and
(\ref{eq:J1}), respectively.

\subsection{Spin expectation values and spin variance}

\begin{figure}
  \centering
  \includegraphics[scale=0.55]{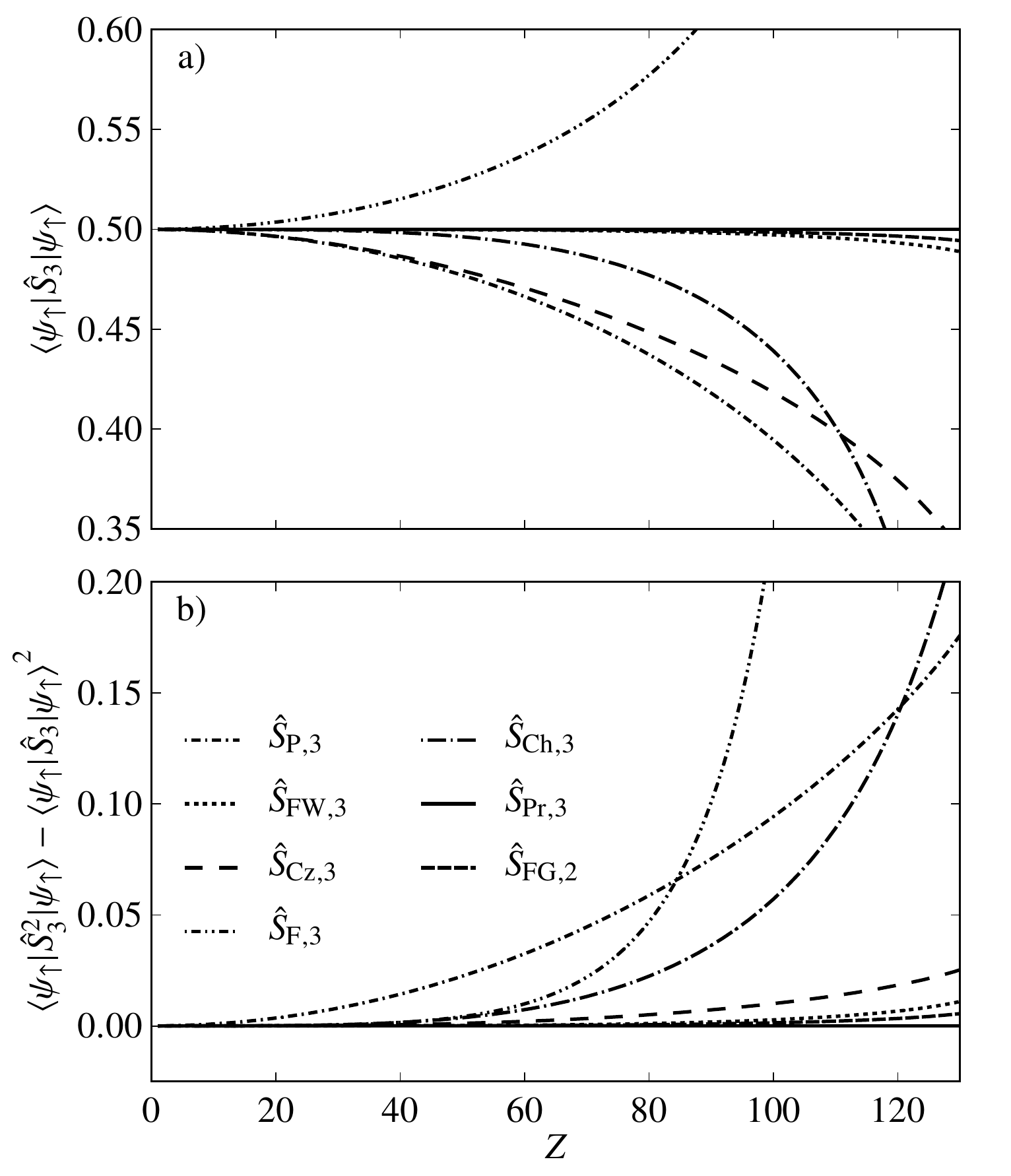}
  \caption{(a) Spin expectation values, adopted from
    \cite{Bauke:etal:2014:What_is}) and (b) spin variance of various
    relativistic spin operators for the hydrogenic ground state
    (\ref{eq:hydrogen_real_1}) as a function of the atomic number $Z$.
    For the ground state (\ref{eq:hydrogen_real_2}) we find the same
    spin variance and the same spin expectation values but with opposite
    sign (not displayed in the plots).}
  \label{fig:hydrogen_spin}
\end{figure}

In momentum space representation, the relativistic spin operators
introduced in Sec.~\ref{sec:spins} are simple matrices, thus, with the
momentum space representation (\ref{eq:hydrogen_momentum}) spin
expectation values as well as the spin variance of the degenerate
hydrogenic ground states can be calculated.  For simplicity, we
calculate spin expectation values in the $z$~direction, that is, $\vec
n=\trans{(0,0,1)}$, for the reminder of this section.  The spin
expectation values $\braket{\psi_{\uparrow} | \opS[3] |
  \psi_{\uparrow} }$ and the spin variance $\braket{\psi_{\uparrow} |
  \opS[3]^2 | \psi_{\uparrow} }- \braket{\psi_{\uparrow} | \opS[3] |
  \psi_{\uparrow} }^2$ are displayed in
Fig.~\ref{fig:hydrogen_spin}(a) for the state
$\psi_{\uparrow}=\psi_{1, 1, \frac{1}{2}, \frac{1}{2}}$ as a function
of the atomic number $Z$.  In general, the spin expectation values and
the spin variance are complicated functions of the nuclear charge $Z$.
For the Pauli and the Pryce spin operators, however, the spin
expectation values and the spin variance are given explicitly by
\begin{subequations}
  \begin{align}
    \braket{\psi_{\uparrow} | \opSP[3] | \psi_{\uparrow} } & =
    \frac{1}{6}\left(1+2\sqrt{1-Z^2\alpha_\mathrm{el}^2}\right)\,, \\
    \braket{\psi_{\uparrow} | \opSP[3]^2 | \psi_{\uparrow} }-
    \braket{\psi_{\uparrow} | \opSP[3] | \psi_{\uparrow} }^2 & =
    \frac{1}{4}-\left( \frac{1}{6}\left(1+2\sqrt{1-Z^2\alpha_\mathrm{el}^2}\right)
    \right)^2
  \end{align}
\end{subequations}
and
\begin{subequations}
  \begin{align}
    \braket{\psi_{\uparrow} | \opSPr[3] | \psi_{\uparrow} } & =
    \frac{1}{2} \,,\\
    \braket{\psi_{\uparrow} | \opSPr[3]^2 | \psi_{\uparrow} }-
    \braket{\psi_{\uparrow} | \opSPr[3] | \psi_{\uparrow} }^2 & =
    0\,,
  \end{align}
\end{subequations} 
respectively.

For small atomic numbers ($Z\lessapprox20$), all spin operators yield
about $1/2$; for larger $Z$ when relativistic effects set in, however,
expectation values differ significantly from each other.  While for
Pauli, Fouldy-Wouthuysen, Czachor, Chakrabarti, and Fradkin-Good spin
operators the spin expectation value is reduced, the expectation value
of the Frenkel spin operator exceeds $1/2$.  Only for the Pryce
operator do we find that the spin expectation values is $1/2$ for all
values of $Z$, implying zero spin variance as shown in
Fig.~\ref{fig:hydrogen_spin}(b). Spin expectation values and spin
variances for the state $\psi_\downarrow=\psi_{1, 1, \frac{1}{2},
  -\frac{1}{2}}$ follow by symmetry via
\begin{equation}
  \braket{\psi_{\uparrow}|\opS[3]|\psi_{\uparrow}}=
  -\braket{\psi_{\downarrow}|\opS[3]|\psi_{\downarrow}}
\end{equation}
and 
\begin{equation}
  \braket{\psi_{\uparrow} | \opS[3]^2 | \psi_{\uparrow} } -
  \braket{\psi_{\uparrow} | \opS[3] | \psi_{\uparrow} }^2 =
  \braket{\psi_{\downarrow} | \opS[3]^2 | \psi_{\downarrow} } -
  \braket{\psi_{\downarrow} | \opS[3] | \psi_{\downarrow} }^2 \,.
\end{equation}

The different predictions for spin expectation values and spin variances
that follow from different definitions of the relativistic spin operator
may serve as a basis for an experimental test for a relativistic spin
operator \cite{Bauke:etal:2014:What_is} that is implemented by a
particular spin measurement experiment.  We assume that the electron of
a highly charged hydrogenlike ion was prepared in its ground state
$\psi_{\uparrow}$, e.\,g., by exposing the ion to a strong magnetic
field in the $z$~direction and turning it off adiabatically.  A spin
measurement experiment for such a state will yield spin $1/2$ with
probability $P_\uparrow=1/2+\braket{\psi_{\uparrow} | \opS[3] |
  \psi_{\uparrow} }$ where $\opS[3]$ is the spin operator that is
realized by the particular measurement procedure.  For hydrogenlike
uranium, $Z=92$, our theoretical predictions yield, for example, for the
Pauli operator $P_\uparrow=91.4\,\%$, for the Fouldy-Wouthuysen operator
$P_\uparrow=99.8\,\%$, and $P_\uparrow=100\,\%$ in the case of the Pryce
operator.  Note that it is a completely open question how experimental
procedures that aim to measure the electron spin state map to
mathematical spin operators.

\subsection{Pryce spin operator in systems with spherical symmetry} 

We demonstrated that only the Pryce spin operator yields spin
expectation values of $\pm1/2$ for the ground states of hydrogenlike
ions.  In the following we will show that this is a consequence of the
spherical symmetry of the Coulomb potential and that each system with
spherical symmetry has some energy eigenstates that are eigenstates of
the Pryce spin operator, too.  On can show
\cite{Greiner:2000:RQM,Bjorken:Drell:1964:RQM} that every bound
eigenstate of any system with spherically symmetric potential $\phi(|\vec r|)$,
\begin{equation}
  \label{eq:Hs}
  \op{H}_{s} = \opHo + q\phi(|\vec r|)\,,
\end{equation}
has the form (\ref{eq:hydrogen_eigen_1}) or (\ref{eq:hydrogen_eigen_2}),
respectively.  Only the radial functions $g_{n,\kappa,j}(r)$ and
$f_{n,\kappa,j}(r)$ depend on the specific potential.  Consequently, the
momentum space representations of \eqref{eq:hydrogen_eigen} have for
$\kappa =j+{1}/{2}$ the form
\begin{subequations}
  \label{eq:hydrogen_eigen_momentum}
  \begin{equation}
    \label{eq:hydrogen_eigen_1_momentum}
    \tilde\psi_{n,\kappa,j,m}(p,\theta',\phi') =
    \begin{pmatrix}
      \tilde g_{n,\kappa,j}(p)\sqrt{\frac{j+m}{2j}}Y_{j-1/2,m-1/2}(\theta', \phi') \\[1.5ex]
      \tilde g_{n,\kappa,j}(p)\sqrt{\frac{j-m}{2j}}Y_{j-1/2,m+1/2}(\theta', \phi') \\[1.5ex]
      -\tilde f_{n,\kappa,j}(p)\I\sqrt{\frac{j-m+1}{2j+2}}Y_{j+1/2,m-1/2}(\theta', \phi') \\[1.5ex]
      \tilde f_{n,\kappa,j}(p)\I\sqrt{\frac{j+m+1/2}{2j+2}}Y_{j+1/2,m+1/2}(\theta', \phi')
    \end{pmatrix}
  \end{equation}
  and for $\kappa =-j-{1}/{2}$
  \begin{equation}
    \label{eq:hydrogen_eigen_2_momentum}
    \tilde\psi_{n,\kappa,j,m}(p,\theta',\phi') =
    \begin{pmatrix}
      -\tilde g_{n,\kappa,j}(p)\sqrt{\frac{j-m+1}{2j+2}}Y_{j+1/2,m-1/2}(\theta', \phi') \\[1.5ex]
      \tilde g_{n,\kappa,j}(p)\sqrt{\frac{j+m+1}{2j+2}}Y_{j+1/2,m+1/2}(\theta', \phi') \\[1.5ex]
      \tilde f_{n,\kappa,j}(p)\I\sqrt{\frac{j+m}{2j}}Y_{j-1/2,m-1/2}(\theta', \phi') \\[1.5ex]
      \tilde f_{n,\kappa,j}(p)\I\sqrt{\frac{j-m}{2j}}Y_{j-1/2,m+1/2}(\theta', \phi')
    \end{pmatrix}\,,
  \end{equation}
\end{subequations}
respectively, with 
\begin{subequations}
  \begin{align}
    \tilde g_{n,\kappa,j}(p) & = 
    \sqrt{\frac{2}{\pi}}(-\I)^{j-1/2} \int_0^\infty g_{n,\kappa,j}(r) j_{j-1/2}(rp)r^2\,\D r \,,\\
    \tilde f_{n,\kappa,j}(p) & = 
    \sqrt{\frac{2}{\pi}}(-\I)^{j+1/2} \int_0^\infty f_{n,\kappa,j}(r) j_{j+1/2}(rp)r^2\,\D r 
  \end{align}
  for \eqref{eq:hydrogen_eigen_1_momentum} and 
  \begin{align}
    \tilde g_{n,\kappa,j}(p) & = 
    \sqrt{\frac{2}{\pi}}(-\I)^{j+1/2} \int_0^\infty g_{n,\kappa,j}(r) j_{j+1/2}(rp)r^2\,\D r \,,\\
    \tilde f_{n,\kappa,j}(p) & = 
    \sqrt{\frac{2}{\pi}}(-\I)^{j-1/2} \int_0^\infty f_{n,\kappa,j}(r) j_{j-1/2}(rp)r^2\,\D r 
  \end{align}
\end{subequations}
for \eqref{eq:hydrogen_eigen_2_momentum} (see also the appendix).  The
momentum space representation of the Pryce spin operator in the
$z$~direction is
\begin{equation}
  \label{eq:Pryce_momentum}
  \opSPr[3] =
  \begin{pmatrix}
    \frac 12 & 0 & 0 & 0 \\[0.5ex] 
    0 & -\frac 12 & 0 & 0 \\[0.5ex]
    0 & 0 & -\frac 12 + \cos^2(\theta') & \cos(\theta')\sin(\theta')\e^{-\I\phi'} \\[0.5ex] 
    0 & 0 & \cos(\theta')\sin(\theta')\e^{\I\phi'} & \frac 12 - \cos^2(\theta')
  \end{pmatrix}\,.
\end{equation}
With this result one can show that
\begin{subequations}
  \label{eq:extremal_Pryce}
  \begin{align}
    \opSPr[3] \psi_{n,\kappa,j, j} & =
    +\frac 12 \psi_{n,\kappa,j,j}\,, \\
    \opSPr[3] \psi_{n,\kappa,j,-j} & = 
    -\frac 12 \psi_{n,\kappa,j,-j} 
  \end{align}
\end{subequations}
by employing (\ref{eq:hydrogen_eigen_1_momentum}) and
(\ref{eq:Pryce_momentum}) and expressing the spherical harmonics in
terms of trigonometric functions.  Thus, eigenstates of central
potentials with extremal quantum number $m=\pm j$ are eigenstates of the
Pryce spin operator with eigenvalue $\pm1/2$.

This has an interesting consequence for the spin of superpositions of
states with $m=\pm j$ as, for example, $\Psi(\vec r, t)=c_1\psi_1(\vec
r, t)+c_2\psi_2(\vec r, t)$ with $\psi_i(\vec r, t) =
\psi_{n,j_i+1/2,j_i,j_i}(\vec r)\e^{-\mathcal{E}(n, j_i+1/2)t}$.  The
spin expectation value of this state is given by
\begin{multline}
  \label{eq:mean_S3}
  \Braket{ \Psi | \opS[3] | \Psi} 
  = 
  |c_1|^2\braket{\psi_{1} | \opS[3] | \psi_{1} } + 
  |c_2|^2\braket{\psi_{2} | \opS[3] | \psi_{2} } \\
  + 2 
  \RE\braket{c_1\psi_{1} | \opS[3] | c_2\psi_{2} }\,.
\end{multline}
As the states $\psi_{1}$ and $\psi_{2}$ have different energies the
mixing term $\braket{c_1\psi_{1} | \opS[3] | c_2\psi_{2} }$ and
therefore the spin expectation value oscillates in time with the
frequency $|\mathcal{E}(n, j_1+1/2)-\mathcal{E}(n, j_2+1/2)|$ unless
$\psi_1(\vec r, t)$ and $\psi_2(\vec r, t)$ are eigenfunctions of the
spin operator $\opS[3]$, as is the case for the Pryce spin operator.
The $|\mathcal{E}(n, j_1+1/2)-\mathcal{E}(n, j_2+1/2)|$ varies over
several orders of magnitude depending on the parameters $n$, $j_1$,
$j_2$, and $Z$ and can be made small by increasing the quantum number
$n$.

The Pryce spin operator allows us to establish a notable
correspondence between the relativistic Dirac theory and the
nonrelativistic Pauli theory of systems with spherical symmetry.  In
the nonrelativistic case, the Pauli Hamiltonian for some spherically
symmetric potential $\phi(|\vec r|)$ and the operator of total angular
momentum are defined as
\begin{equation}
  \label{eq:Hs_nr}
  \op{H}_{s,\mathrm{nr}} =
  \frac{\opp^2}{2m_0} + q\phi(|\vec r|)
\end{equation}
and $\opJ_\mathrm{nr} = \vec r \times \opp + \vec\sigma/2$, respectively.
The second term in the nonrelativistic total angular momentum equals the
nonrelativistic Pauli spin operator
\begin{equation}
  \opSPnr = \frac{1}{2}\vec\sigma \,.
\end{equation}
In analogy to the Dirac theory, the two-component eigenfunctions of the
Pauli Hamiltonian \eqref{eq:Hs_nr} can be expressed as simultaneous
eigenstates $\psi^{\mathrm{nr}}_{n,\kappa,j,m}$ of
$\op{H}_{s,\mathrm{nr}}$, $\opJ_\mathrm{nr}^2$, $\opJ[\mathrm{nr},3]$,
and the nonrelativistic spin-orbit operator
$\op{K}_\mathrm{nr}=\mbox{$\vec\sigma\cdot[\vec
  r\times(-\I\vec\nabla)+\mathbb{I}_2)]\}$}$ fulfilling the eigenequations
\cite{Bethe:Salpeter:2008:Atoms,Thaller:2005:vis_QM_II}
\begin{subequations}
  \begin{align}
    \op{H}_{s,\mathrm{nr}} \psi^{\mathrm{nr}}_{n,\kappa,j,m} & = \mathcal{E}_{\mathrm{nr}}(n)\psi^{\mathrm{nr}}_{n,\kappa,j,m} \,, &
    n & = 1,2,\dots\,,\\
    \op{K}_\mathrm{nr} \psi^{\mathrm{nr}}_{n,\kappa,j,m} & = \kappa\psi^{\mathrm{nr}}_{n,\kappa,j,m} \,, &
    |\kappa| & = 1,2,\dots,n \,,\kappa\ne-n\,,\\
    \opJ_\mathrm{nr}^2 \psi^{\mathrm{nr}}_{n,\kappa,j,m} & = j(j+1)\psi^{\mathrm{nr}}_{n,\kappa,j,m} \,, &
    j & = |\kappa|-\tfrac{1}{2} \,,\\
    \opJ[\mathrm{nr},3] \psi^{\mathrm{nr}}_{n,\kappa,j,m} & = m\psi^{\mathrm{nr}}_{n,\kappa,j,m} \,, &
    m & = -j,(j-1),\dots,j \,,
  \end{align}
\end{subequations}
where $\mathcal{E}_\mathrm{nr}(n)$ denotes the eigenenergies.  The
$\psi^{\mathrm{nr}}_{n,\kappa,j,m}$ are in general not eigenfunctions of
$\opSPnr$.  On can show \cite{Bethe:Salpeter:2008:Atoms}, however, that 
\begin{subequations}
  \label{eq:extremal_Pauli}
  \begin{align}
    \opSPnr[3] \psi^{\mathrm{nr}}_{n,\kappa,j, j} & =
    +\frac 12 \psi^{\mathrm{nr}}_{n,\kappa,j,j}\,, \\
    \opSPnr[3] \psi^{\mathrm{nr}}_{n,\kappa,j,-j} & = 
    -\frac 12 \psi^{\mathrm{nr}}_{n,\kappa,j,-j}
  \end{align}
\end{subequations}
hold.  A comparison of \eqref{eq:extremal_Pryce} and
\eqref{eq:extremal_Pauli} shows that the nonrelativistic Pauli spin
operator $\opSPnr$ and the relativistic Pryce spin operator $\opSPr$
play an analogous role within the theories they belong to.

\section{Discussion and conclusions}
\label{sec:conclusions}

We have reconsidered the electron's spin degree of freedom within
relativistic quantum mechanics.  The motivation of our investigation was
the observation that there is no universally accepted spin operator in
the Dirac theory.  In fact, several relativistic spin operators have
been proposed in the literature.  We investigated the properties of some
popular proposed spin operators and analyzed how the different spin
operators can lead to different theoretical predictions for expectation
values of the spin in different physical setups.

The two pairs given by the Pauli and the Czachor spins and by the
Foldy-Wouthuysen and the Chakrabarti spins, respectively, act as
identical operators in each of two the subspaces of free-particle
states with positive and negative energy.  On the basis of the spin
operators' mathematical properties, the Foldy-Wouthuysen and the Pryce
spin operators seem to be the most promising candidates for a proper
relativistic spin operator.  Both operators commute with the free
Dirac Hamiltonian as well as with the total angular momentum operator
and the linear momentum operator.  Furthermore, they obey the angular
momentum algebra and have eigenvalues $\pm1/2$.  The Foldy-Wouthuysen
and the Pryce spin operators are equivalent on the subspace of wave
functions that are superpositions of free-particle eigenstates with
positive energy.  However, we demonstrated in three different physical
setups that if interaction potentials are present one can distinguish
between both operators because they may lead to different expectation
values for the same quantum state.  The three setups reveal a rather
consistent
behavior.  % The Pryce spin operator turned out to be robust in the sense
% that it leads to the least variation as a function of the system's
% parameters as well as of time.

The various proposed spin operators are usually motivated by abstract
theoretical considerations rather than experimental evidence.  The
fact that these spin operators yield different predictions about the
expectation value of the spin in several setups as, for example, for
electrons in scattering at a step potential, electrons in standing
laser fields, or hydrogenic eigenstates as considered here, offers the
opportunity to discriminate between the various proposed spin
operators.  In this way one may rule out some operators for which the
theoretical predictions are incompatible with experimental results.
The identification of the correct relativistic spin operator would
immediately induce relativistic operators for the orbital angular
momentum and the position.  Thus, the identification of the right
description of the spin within the Dirac theory has broad implications
beyond the spin itself.

We provided precise predictions about what could be measured if a spin
measurement procedure implements a physical realization of a
particular spin operator.  However, we did not dwell on how to measure
the spin.  In fact, experiments that measure the spin (and not mere
spin effects) are challenging from a technological point of view as
well as conceptually even today.  As pointed out earlier, it is a
completely open question how experimental measuring procedures map to
mathematical spin operators.  See \cite{Morrison:2007:Spin} for an
in-depth discussion.  There is a ongoing effort to advance spin
measurement techniques.  A Stern-Gerlach experiment for electrons may
be feasible
\cite{Batelaan:etal:1997:Stern-Gerlach,Garraway:Stenholm:1999:Observing_spin,McGregor:etal:2011:Stern-Gerlach}.
The spin may be measured indirectly via transferring it to orbital
angular momentum \cite{Karimi:etal:2012:Spin-to-Orbital}.
\vspace*{1ex plus 8ex}

\begin{acknowledgments}
  We enjoyed several helpful discussions with C.~C.~Gerry,
  K.~Z.~Hatsagortsyan, S.~Meuren, C.~M{\"u}ller, O.~Skoromnik, Q.~Su,
  and E.~Yakaboylu.  R.~G.\ acknowledges the kind hospitality of the
  Max Planck Institute for Nuclear Physics in Heidelberg.  This work
  was supported by the NSF.
\end{acknowledgments}

\appendix*

\makeatletter
\renewenvironment{widetext@grid}{%
  \par\ignorespaces
  \setbox\widetext@top\vbox{%
    % \hb@xt@\hsize{%
    % \leaders\hrule\hfil
    % \vrule\@height6\p@
    %}%
  }%
  \setbox\widetext@bot\hb@xt@\hsize{%
    \vrule\@depth6\p@
    \leaders\hrule\hfil
  }%
  \onecolumngrid
  \vskip10\p@
  \dimen@\ht\widetext@top\advance\dimen@\dp\widetext@top
  \cleaders\box\widetext@top\vskip\dimen@
  \vskip6\p@
  \prep@math@patch
}{%
  \par
  \vskip6\p@
  \setbox\widetext@bot\vbox{%
  % \hb@xt@\hsize{\hfil\box\widetext@bot}%
  }%
  \dimen@\ht\widetext@bot\advance\dimen@\dp\widetext@bot
  \cleaders\box\widetext@bot\vskip\dimen@
  %\vskip8.5\p@
  \twocolumngrid\global\@ignoretrue
  \@endpetrue
}%
\let@environment{widetext}{widetext@grid}%
\makeatother

\begin{widetext}

\section{Fourier transform in spherical coordinates}
\label{sec:Fourier}

Using the vectors $\vec r$ and $\vec p$ and their representation in
spherical coordinates %
$\vec r=\trans{(r\sin\theta\cos\phi, r\sin\theta\sin\phi,
  r\cos\theta)}$ and %
$\vec p=\trans{(p\sin\theta'\cos\phi', r'\sin\theta'\sin\phi',
  r'\cos\theta')}$ the (inverse) Fourier transform of some function
$f(\vec r)$ is defined as
\begin{align}
  \nonumber
  \mathcal{F}^\pm[f(\vec r)] & =
  \frac{1}{(2\pi)^{3/2}} \int f(\vec r) \exp(\pm\I\,\vec r\cdot\vec p)\,\D^3r \\
  & = \frac{1}{(2\pi)^{3/2}}
  \int_0^\infty\!\!\int_0^\pi\!\!\int_0^{2\pi}
  f(r, \theta, \phi) \exp\left(
    \pm\I rp\left(
      \sin\theta\sin\theta'\cos(\phi-\phi')+
      \cos\theta\cos\theta'
    \right)
  \right)r^2\sin\theta
  \,\D\phi\,\D\theta\,\D r\,. \\
  \intertext{For functions not depending on the azimuthal angle $f(r,
    \theta, \phi)=f(r, \theta)$ the (inverse) Fourier transform simplifies to} 
  \mathcal{F}^\pm[f(\vec r)] & = 
  \frac{1}{\sqrt{2\pi}}
  \int_0^\infty\!\!\int_0^\pi
  f(r, \theta) J_0(rp\sin\theta\sin\theta')
  \exp\left(\pm \I rp\cos\theta\cos\theta'\right)r^2\sin\theta
  \,\D\theta\,\D r
  \intertext{with $J_0(x)$ denoting the zeroth-order Bessel function of the
    first kind.  For spherically symmetric functions $f(r, \theta,
    \phi)=f(r)$ we finally get} 
  \mathcal{F}^\pm[f(\vec r)] & = 
  \sqrt{\frac{2}{\pi}}\int_0^\infty
  f(r) r^2 j_0(rp)\,\D r\,.
\end{align}
Note that for spherically symmetric functions the Fourier transform and
its inverse have the same form.\pagebreak[1]

If the radial and angular dependences separate, e.\,g., $f(r, \theta,
\phi)=R(r)Y_{l,m}(\theta, \phi)$ with $Y_{l,m}(\theta, \phi)$ denoting
the complex-valued orthonormal spherical harmonics, it is beneficial to
utilize the generalized Jacobi-Anger identities
\begin{equation}
  \label{eq:Jacobi-Anger}
  \frac{1}{(2\pi)^{3/2}} \exp(\pm\I\,\vec r\cdot\vec p) =
  \sqrt{\frac{2}{\pi}} \sum_{l=0}^\infty (\pm\I)^l j_l(rp)
  \sum_{m=-l}^l Y_{l,m}(\theta, \phi)Y_{l,m}^*(\theta', \phi') 
  =
  % \label{eq:Jacobi-Anger}
  \sqrt{\frac{2}{\pi}} \sum_{l=0}^\infty (\pm\I)^l j_l(rp) 
  \sum_{m=-l}^l Y_{l,m}^*(\theta, \phi)Y_{l,m}(\theta', \phi')
\end{equation}
with the spherical Bessel functions of the first kind $j_l(x)$ as
defined in \cite{Arfken:Weber:2005:Mathematical_Methods}.  By using
(\ref{eq:Jacobi-Anger}) and the orthonormalization of the spherical
harmonics, we find
\begin{equation}
  \mathcal{F}^\pm[R(r)Y_{l,m}(\theta, \phi)] =
  (\pm\I)^l Y_{l,m}(\theta', \phi')\sqrt{\frac{2}{\pi}}
  \int_0^\infty R(r) j_l(rp)r^2\,\D r\,.
\end{equation}
Specifying $R(r)$ as $R(r)={\e^{-zr}}/{(2zr)^{1-\gamma}}$, we may
define the two functions $\mathcal{J}_0 (z, \gamma, p )$ and
$\mathcal{J}_1 (z, \gamma, p )$ as 
\begin{align}
  \label{eq:J0}
  \mathcal{J}_0 (z, \gamma, p ) = 
  \sqrt{\frac{2}{\pi}}
  \int_0^\infty\frac{\e^{-zr}}{(2zr)^{1-\gamma}}r^2 j_0(rp)\,\D r & =
  \sqrt{\frac{2}{\pi}}
  \frac{2^{\gamma-1}\GAMMA(\gamma+1)}
  {z^2p(1+p^2/z^2)^{(\gamma+1)/2}} \sin((1+\gamma)\arctan(p/z)) \,, \\
  \nonumber
  \mathcal{J}_1 (z, \gamma, p ) = 
  \sqrt{\frac{2}{\pi}}
  \int_0^\infty\frac{\e^{-zr}}{(2zr)^{1-\gamma}}r^2 j_1(rp)\,\D r & \\
  \label{eq:J1}
  & \hspace{-3cm}=\sqrt{\frac{2}{\pi}}
  \frac{2^{\gamma-1}\GAMMA(\gamma)}
  {z^2p(1+p^2/z^2)^{(\gamma+1)/2}}
  (
  -(1+\gamma)\cos((1+\gamma)\arctan(p/z)) + \frac{z}{p}\sin((1+\gamma)\arctan(p/z))
  )
\end{align}
for $0<\gamma<1$. The Eqs.~(\ref{eq:J0}) and~(\ref{eq:J1}) are two
special cases of a more general formula that may be useful in
investigating excited states of the hydrogen atom (see Eq.~(32.7) in
\cite{Sneddon:1963:Spezielle_Funktionen}).

\end{widetext}

\bibliography{relativistic_spin}

\end{document}